\documentclass[a4paper,fleqn]{cas-dc}
\AtBeginDocument{\setlength\mathindent{0pt}}
\usepackage[numbers]{natbib}

\usepackage{mathtools}
\usepackage{bm}

\usepackage{graphicx}
\usepackage{subfig}

\usepackage{algorithm}
\usepackage{algorithmic}

\usepackage[expansion=false]{microtype}
\usepackage{enumitem}

\definecolor{darkred}{rgb}{.7,0,0}

\definecolor{darkgreen}{rgb}{.15,.55,0}

\definecolor{darkblue}{rgb}{0,0,0.7}

\newcommand{\design}{\xi}
\newcommand{\designset}{\Xi}

\newcommand{\param}{\theta}
\newcommand{\Param}{\Theta}
\newcommand{\paramset}{\bm{\Theta}}

\newcommand{\Alpha}{\mathcal{A}}
\newcommand{\Beta}{\mathcal{B}}
\newcommand{\Epsilon}{\mathcal{E}}

\newcommand{\argmax}{\operatornamewithlimits{argmax}}

\begin{document}
\let\WriteBookmarks\relax
\def\floatpagepagefraction{1}
\def\textpagefraction{.001}

\shorttitle{Variational GO-OED for Mixed-Distribution QoIs}
\shortauthors{C. Cheng, X. Huan and Y. Pan}

\title[mode=title]{Variational Goal-Oriented Optimal Experimental Design for Mixed-Distribution Quantities of Interest: Application to Ship Roll Safety}

\author[1]{Chen Cheng}[orcid=0000-0001-9773-2993]
\ead{c.cheng.6202@gmail.com}
\credit{Methodology, Software, Formal analysis, Writing -- original draft}

\author[1]{Xun Huan}[orcid=0000-0001-6544-2764]
\ead{xhuan@umich.edu}
\ead[url]{https://uq.engin.umich.edu/}
\cormark[1]
\credit{Conceptualization, Methodology, Supervision, Writing -- review \& editing}

\author[2]{Yulin Pan}[orcid=0000-0002-7504-8645]
\ead{yulinpan@umich.edu}
\ead[url]{https://sites.google.com/umich.edu/fpel}
\cormark[1]
\credit{Conceptualization, Methodology, Supervision, Writing -- review \& editing}

\affiliation[1]{
organization={Department of Mechanical Engineering, University of Michigan},
city={Ann Arbor},
postcode={48109},
state={MI},
country={USA}}

\affiliation[2]{
organization={Department of Naval Architecture and Marine Engineering, University of Michigan},
city={Ann Arbor},
postcode={48109},
state={MI},
country={USA}}

\cortext[1]{Corresponding author}

\begin{abstract}
Goal-oriented optimal experimental design (GO-OED) selects experiments according to the expected information gain (EIG) about a quantity of interest (QoI) rather than the full parameter vector. This work develops a variational GO-OED formulation for mixed discrete--continuous QoI laws arising in probabilistic mechanics when thresholding or event-based transformations map a positive-probability set of uncertain inputs to a common value while other inputs produce continuously varying responses. The motivating application is ship roll safety assessment in random waves, where the QoI is the temporal exceedance probability above a prescribed roll-angle threshold. This quantity is zero when no exceedance occurs and varies continuously over positive values otherwise. A purely continuous variational approximation does not dominate a posterior QoI law containing an atom, yielding an infinite Kullback--Leibler divergence and a trivial Barber--Agakov lower bound of $-\infty$. Scoring atom samples using continuous density values instead changes the objective and does not produce a valid lower-bound estimator. We introduce a mixed variational approximation that models the conditional atom probability and continuous component separately, with a normalizing flow used for the latter. An analytical example recovers the correct EIG landscape, while the ship roll application provides stable EIG lower-bound estimates and identifies informative wave conditions for temporal-exceedance-probability inference.
\end{abstract}

\begin{highlights}
\item A mixed-measure variational GO-OED formulation is developed for QoIs with atoms.

\item Purely continuous variational families yield an infinite KL gap when an atom is present.

\item An atom-plus-continuous normalizing-flow model provides a finite EIG lower bound.

\item An analytical example exposes the failure of density scoring at an atom.

\item Ship-wave designs are optimized for inference of temporal exceedance probability.
\end{highlights}

\begin{keywords}
Expected information gain \sep
Normalizing flows \sep
Barber--Agakov lower bound \sep
Temporal exceedance probability \sep
Threshold-based inference \sep
Variational inference \sep
Uncertainty quantification
\end{keywords}

\maketitle

\section{Introduction}
\label{sec:intro}

Many experiments in probabilistic mechanics are conducted not to identify every uncertain model parameter, but to reduce uncertainty in a downstream prediction, safety metric, or decision-relevant quantity of interest (QoI).
This distinction is especially important when the QoI is defined through thresholding, censoring, or event occurrence, as in exceedance duration, failure severity, accumulated damage beyond onset, or the fraction of time spent in an unsafe state.

\begin{figure*}
    \centering
    \includegraphics[width=\linewidth]{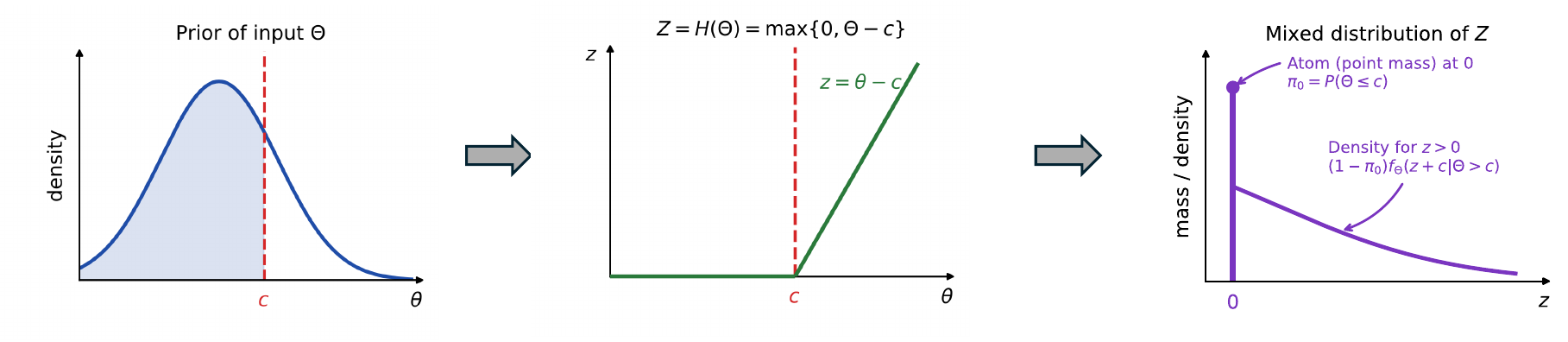}
    \caption{
    Illustration of a threshold-induced mixed QoI law.
    A continuously distributed parameter $\Theta$ is mapped through
    $Z=H(\Theta)=\max\{0,\Theta-c\}$.
    The positive-probability region $\Theta\le c$ is mapped to the common value
    $Z=0$, producing an atom with probability
    $\pi_0=\mathbb{P}(\Theta\le c)$.
    For $\Theta>c$, the transformation is one-to-one and yields the continuous component on $z>0$, whose shape is the right tail of the density of $\Theta$ shifted left by $c$.
    }
    \label{fig:mixed-qoi-illustration}
\end{figure*}

Such transformations can induce mixed QoI laws even when the underlying uncertain parameters are continuously distributed.
As illustrated in Figure~\ref{fig:mixed-qoi-illustration}, a positive-probability subset of parameter space may be mapped to a common QoI value, while the remaining parameter values produce continuously varying outcomes.
The resulting law contains a probability atom together with a continuous component.
This mixed structure is a defining feature of the QoI distribution and must be preserved in uncertainty representation and inference.
Candidate experiments may differ substantially in how much they reduce uncertainty in this downstream quantity, motivating an experimental-design criterion that evaluates their anticipated effect on the QoI directly.

Bayesian optimal experimental design (OED) provides a principled framework for selecting experiments under uncertainty~\cite{Chaloner1995,Ryan2016,Alexanderian2021,Rainforth2024,Strutz2024,Huan2024}.
A statistical or physics-based model predicts the observations that may arise under each candidate design, allowing experiments to be compared before data are collected.
A common Bayesian design criterion is the expected information gain (EIG), equivalently the mutual information between an inferential target and a future observation~\cite{Lindley1956}.
When the inferential target is the uncertain parameter vector, we refer to the resulting formulation as parameter-oriented OED (PO-OED), which underlies most conventional Bayesian OED approaches.

Parameter inference is not always the final purpose of an experiment.
In mechanics applications, uncertain parameters often matter primarily through a structural response, reliability metric, rare-event probability, or safety-relevant prediction.
When the parameter-to-QoI map is many-to-one, experiments that distinguish parameter configurations may provide little information about the downstream quantity.
Goal-oriented OED (GO-OED) addresses this mismatch by maximizing the EIG in the QoI rather than in the full parameter vector~\cite{Bernardo1979,Attia2018,Wu2023,Chakraborty2024,Shen2025,Zhong2026}.
The resulting criterion favors observations that are expected to reduce uncertainty in the QoI directly while allowing uncertainty to remain in parameter directions that do not affect it.
This alignment is particularly relevant in probabilistic safety and reliability analysis, where the inferential target is commonly a threshold-based response or risk metric.

Computational methods for nonlinear GO-OED have largely represented prior and posterior QoI uncertainty using continuous densities or density ratios~\cite{Chakraborty2024,Shen2025,Zhong2026}.
These representations are appropriate when the QoI law is absolutely continuous with respect to Lebesgue measure, but they do not extend directly to threshold-induced mixed distributions.
If a positive-probability subset of parameter space is mapped to a common QoI value $z_0$, the QoI law contains an atom at $z_0$ together with a continuous remainder.
Its natural dominating measure therefore combines a Dirac mass at $z_0$ with Lebesgue measure on the continuous portion of the QoI space.

This distinction is important for variational GO-OED.
One class of variational estimators introduces a parameterized approximation to the posterior law of the inferential target and optimizes a Barber--Agakov lower bound~\cite{Barber2003,Foster2019,Dong2025,Shen2025}.
The corresponding Kullback--Leibler (KL) gap is finite only when the true posterior law is absolutely continuous with respect to the variational approximation.
A purely continuous variational law assigns zero probability to every singleton, including the atom location $z_0$.
Whenever the posterior QoI law assigns positive probability to that outcome, the domination condition fails and the KL divergence is infinite.
The correctly defined Barber--Agakov objective then equals $-\infty$ and remains only a trivial, unusable lower bound.

The same measure mismatch appears in finite implementations when atom samples are scored using continuous density values rather than probability masses.
The density value at the numerical location of an atom is not the probability assigned to that singleton.
Replacing the correct zero mixed-measure derivative with a continuous density value changes the objective, producing an empirical criterion that is not a valid Barber--Agakov lower-bound estimator and may exceed the true EIG.
This failure is structural and cannot be corrected by increasing the expressiveness of the continuous density family or the sample size.

We develop a mixed-distribution variational formulation of GO-OED that resolves this incompatibility.
The proposed approximation represents the atom at $z_0$ through a conditional probability and the continuous component through a conditional normalizing-flow (NF) density~\cite{Rezende2015,Papamakarios2021,Kobyzev2021}.
Both components are defined with respect to the same mixed dominating measure as the true QoI law.
Samples at the atom are scored using log probability mass, while samples in the continuous component are scored using log density.
Under the stated support conditions, this construction yields a finite Barber--Agakov lower bound with a well-defined KL gap.

The methodology is demonstrated through ship roll safety assessment in random waves.
Large-amplitude rolling and capsizing are among the most severe hazards in maritime operations~\cite{Thompson1997,Spyrou2000,Belenky1993}.
Uncertain parameters in nonlinear roll models describe damping, restoring behavior, and wave excitation, and they must often be inferred from wave-basin experiments or full-scale measurements~\cite{Roberts1986,RobertsVasta2000,Dostal2012}.
The applied wave condition determines which aspects of the stochastic dynamics are revealed, so experimental design is important for resolving safety-relevant behavior efficiently.

The QoI is the temporal exceedance probability, defined as the fraction of an assessment horizon during which the absolute roll angle exceeds a prescribed threshold.
Unlike a group-maximum statistic, it accounts for exceedance duration and aggregates the response over the stochastic wave environment~\cite{Gong2022}.
The map from roll-model parameters to this QoI is nonlinear and many-to-one, so distinct parameter combinations may have similar safety implications.
GO-OED therefore targets information about the temporal exceedance probability directly rather than requiring information about all model parameters.

The temporal exceedance probability also exhibits the mixed-distribution structure central to this work.
The physical roll-angle threshold $r_s$ defines exceedance through $|r(t;\param)|>r_s$, whereas the atom location $z_0$ lies in the QoI space.
Because the QoI is the fraction of time spent above $r_s$, it equals zero when no exceedance occurs, so $z_0=0$.
When exceedance occurs, the QoI takes positive, continuously varying values.
If the non-exceeding parameter regime has positive prior probability, the induced QoI law contains an atom at zero together with a positive continuous component.

A wave-group representation of the target narrow-band sea state provides a tractable design space~\cite{Gong2022}.
Localized wave events are parameterized by their half-length and peak amplitude, yielding a two-dimensional set of physically realizable experimental conditions.
The full stochastic sea state defines the safety QoI, whereas a single localized wave group defines a candidate experiment.
A surrogate auxiliary-response model enables repeated evaluation of the nonlinear roll dynamics within the variational GO-OED computation.

The main contributions of this work are as follows.
\begin{itemize}[noitemsep]
\item
We formulate a mixed-measure variational representation of goal-oriented EIG for QoI laws containing an atom and a continuous component.

\item
We show that a purely continuous variational approximation violates the domination condition whenever the posterior QoI law contains an atom, yielding an infinite KL gap and a trivial Barber--Agakov lower bound of $-\infty$.

\item
We introduce a mixed variational approximation that combines a conditional atom-probability model with a conditional NF for the continuous component, yielding a finite lower-bound objective under the stated support assumptions.

\item
We validate the formulation against an exact goal-oriented EIG in an analytical benchmark and demonstrate its use for nonlinear ship roll safety assessment over a physically defined wave-group design space with surrogate-assisted response evaluation.
\end{itemize}

The remainder of this paper is organized as follows.
Section~\ref{s:BOED} presents the Bayesian GO-OED formulation.
Section~\ref{s:methods} develops the mixed-distribution variational lower bound and the mixed NF approximation.
Section~\ref{s:toy} validates the formulation using an analytically tractable benchmark.
Section~\ref{sec:ship} presents the ship roll safety application.
Section~\ref{s:conclusion} concludes the paper.
\section{Bayesian Goal-Oriented Optimal Experimental Design}
\label{s:BOED}

Consider an uncertain parameter vector\footnote{We use uppercase symbols for random variables and lowercase symbols for their realizations. For example, the random variables $\Param$ and $Y$ take realizations $\param$ and $y$, respectively.}
$\Param\in\paramset\subseteq\mathbb{R}^{p}$, an experimental design $\design\in\designset\subseteq\mathbb{R}^{d}$, and an observation $Y\in\mathcal{Y}\subseteq\mathbb{R}^{n}$.
An experiment consists of selecting $\design$, conducting the experiment under the prescribed conditions, and observing a realization $y$.
The design affects the data-generating mechanism and therefore determines what can be learned from the resulting observation.
Bayesian OED formalizes the selection of $\design$ through an expected utility that quantifies the anticipated statistical value of the experiment.
We consider batch design, in which all experimental conditions are selected before any data are observed.
Sequential extensions that adapt subsequent designs to previously collected observations are left for future work.
Further background on Bayesian OED and predictive GO-OED can be found in~\cite{Huan2024,Zhong2026}.

\subsection{Bayesian inference}
\label{ss:bayes_update}

Prior uncertainty about $\Param$ is represented by the density $p(\param)$.\footnote{We omit random-variable subscripts from probability density notation when the corresponding variables are clear from their arguments. Subscripts are retained for probability laws and whenever needed to avoid ambiguity.}
For an experiment conducted at design $\design$, the likelihood of observing $Y=y$ given $\Param=\param$ is $p(y|\param,\design)$.
After observing $y$, Bayes' rule gives
\begin{align}
    p(\param|y,\design)
    =
    \frac{
        p(y|\param,\design)p(\param)
    }{
        p(y|\design)
    },
    \label{e:bayes}
\end{align}
where the prior is assumed to be independent of the selected design and
\begin{align}
    p(y|\design)
    =
    \int
    p(y|\param,\design)
    p(\param)
    \,\mathrm{d}\param
    \label{e:evidence}
\end{align}
is the prior predictive density, also called the marginal likelihood or model evidence.

The likelihood is often induced by an observation model of the form
\begin{align}
    Y
    =
    G(\Param,\design)+\Epsilon,
    \label{e:observation_model_general}
\end{align}
where
\begin{align}
    G:\paramset\times\designset\rightarrow\mathcal{Y}
\end{align}
is the parameter-to-observation map and $\Epsilon$ represents observational noise.
Different designs alter the relationship between $\Param$ and $Y$ and therefore induce different posterior distributions.

\subsection{Expected information gain}
\label{ss:eig}

Let $Q$ denote the inferential target, which may be the parameter $\Param$ itself or a derived quantity of interest.
For an observation $Y=y$ obtained under design $\design$, we define the realized utility as the information gained about $Q$:
\begin{align}
    u_Q(y,\design)
    =
    D_{\mathrm{KL}}
    \left(
        \mathbb{P}_{Q|Y,\design}(\cdot|y,\design)
        \,||\,
        \mathbb{P}_Q
    \right),
    \label{e:realized_information_gain}
\end{align}
where $\mathbb{P}_Q$ and $\mathbb{P}_{Q|Y,\design}(\cdot|y,\design)$ denote the prior and posterior probability laws of $Q$, respectively.
This measure-level definition accommodates discrete, continuous, and mixed distributions.
Because $u_Q(y,\design)$ depends on the realized observation, its value is not known before the experiment is performed.

The expected utility averages the realized utility over the observations that may arise under design $\design$:
\begin{align}
    U_Q(\design)
    &=
    \mathbb{E}_{Y|\design}
    \left[
        u_Q(Y,\design)
    \right]
    \nonumber\\
    &=
    \mathbb{E}_{Y|\design}
    \left[
        D_{\mathrm{KL}}
        \left(
            \mathbb{P}_{Q|Y,\design}(\cdot|Y,\design)
            \,||\,
            \mathbb{P}_Q
        \right)
    \right]
    \nonumber\\
    &=
    \mathcal{I}(Q;Y|\design),
    \label{e:EIG_general}
\end{align}
where $\mathcal{I}(Q;Y|\design)$ is the mutual information between $Q$ and $Y$ under design $\design$.
Thus, when the realized utility is defined by information gain, the expected utility is the EIG.
The corresponding OED problem is
\begin{align}
    \design^\ast
    \in
    \argmax_{\design\in\designset}
    U_Q(\design).
    \label{e:optimal_design_general}
\end{align}

When the inferential target is the parameter vector, $Q=\Param$, the expected utility becomes
\begin{align}
    U_{\mathrm{PO}}(\design)
    =
    \mathcal{I}(\Param;Y|\design),
    \label{e:PO_EIG}
\end{align}
which defines what we refer to here as PO-OED.
This criterion selects experiments that are expected to reduce uncertainty in the full parameter vector.

\subsection{Goal-oriented optimal experimental design}
\label{ss:GO_OED}

In many applications, the parameters are relevant primarily through a downstream quantity of interest
\begin{align}
    Z
    =
    H(\Param),
    \label{e:goal_map}
\end{align}
where
\begin{align}
    H:\paramset\rightarrow\mathcal{Z}
\end{align}
is a deterministic parameter-to-QoI map.
GO-OED instead uses the EIG about $Z$:
\begin{align}
    U_{\mathrm{GO}}(\design)
    =
    \mathcal{I}(Z;Y|\design).
    \label{e:GO_EIG}
\end{align}
This criterion prioritizes observations that are informative about the downstream QoI rather than requiring uncertainty reduction throughout the full parameter vector.

Because $Z$ is a deterministic function of $\Param$, the variables satisfy the conditional Markov structure
\begin{align}
    Z
    \longleftarrow
    \Param
    \longrightarrow
    Y
    \qquad
    \text{given }\design.
\end{align}
The data-processing inequality therefore gives
\begin{align}
    U_{\mathrm{GO}}(\design)
    =
    \mathcal{I}(Z;Y|\design)
    \leq
    \mathcal{I}(\Param;Y|\design)
    =
    U_{\mathrm{PO}}(\design).
    \label{e:DPI}
\end{align}
Equality holds when $Z$ retains all information in $\Param$ that is relevant to $Y$, equivalently when
\begin{align}
    \mathcal{I}(\Param;Y|Z,\design)
    =
    0.
    \label{e:DPI_equality}
\end{align}
Injectivity of $H$ is sufficient for equality, but it is not necessary.

When $H$ is many-to-one, distinct parameter values may produce the same QoI.
PO-OED may then favor experiments that distinguish parameter configurations with different parameter values but equivalent downstream implications.
GO-OED instead concentrates experimental effort on distinctions that affect the QoI itself.
This distinction is particularly important in safety, reliability, and decision-oriented applications, where the inferential target is often a derived risk or performance metric rather than the complete parameter vector.
\section{Mixed-Distribution Variational GO-OED}
\label{s:methods}

Our computational aim is to evaluate and optimize the GO-OED objective
$U_{\mathrm{GO}}(\design)$ when the QoI
\begin{align}
    Z
    =
    H(\Param)
\end{align}
has a mixed discrete--continuous distribution.
For a fixed design $\design$ and realized observation $y$, the posterior law of $Z$ is obtained by pushing
$\mathbb{P}_{\Param|Y,\design}(\cdot|y,\design)$ through $H$.
The prior law $\mathbb{P}_Z$ is obtained analogously by pushing $\mathbb{P}_{\Param}$ through $H$.
Direct evaluation of $U_{\mathrm{GO}}(\design)$ generally requires these pushforward laws to be approximated for many possible observations and candidate designs.

A sampling-based approach can approximate the posterior pushforward by drawing samples from
$\mathbb{P}_{\Param|Y,\design}(\cdot|y,\design)$ and evaluating $H$ at the resulting parameter samples~\cite{Zhong2026}.
Although conceptually direct, this approach can be prohibitively expensive within design optimization because posterior sampling and goal-space distribution estimation must be repeated across observations and designs.
We instead adopt a variational formulation~\cite{Foster2019,Dong2025,Shen2025} that approximates the conditional law of $Z$ and yields a tractable lower bound on the goal-oriented EIG.

For a mixed QoI, the variational family must preserve the underlying measure structure.
Outcomes at an atom must be assigned probability mass, whereas outcomes in the continuous component must be assigned density with respect to Lebesgue measure.
We first describe the mixed prior and posterior laws, then introduce a common dominating measure that provides a unified pointwise score for their discrete and continuous components.

\subsection{Mixed prior and posterior laws}
\label{ss:mixed_dist}

We focus on the case in which the QoI law contains one atom at $z_0$ and is otherwise absolutely continuous with respect to Lebesgue measure.
Suppose that a subset of parameter space with positive prior probability is mapped to a common QoI value $z_0$.
Define its preimage as
\begin{align}
    \paramset_0
    =
    \left\{
        \param\in\paramset
        |
        H(\param)=z_0
    \right\},
    \label{e:atom_preimage}
\end{align}
and let
\begin{align}
    \pi_0
    =
    \mathbb{P}(\Param\in\paramset_0)
    =
    \mathbb{P}(Z=z_0)
    >
    0
    \label{e:pi0_def}
\end{align}
denote the prior probability of the atom.

We assume that the conditional law of $Z$ given $Z\neq z_0$ is absolutely continuous with respect to Lebesgue measure $\mu$ on a support
$\mathcal{Z}_{\mathrm{c}}\subseteq\mathcal{Z}\setminus\{z_0\}$.
Let $f_{\mathrm{c}}$ denote its normalized conditional density, so that
\begin{align}
    \int
    f_{\mathrm{c}}(z)
    \,\mathrm{d}z
    =
    1.
\end{align}
The prior law of $Z$ is therefore
\begin{align}
    \mathbb{P}_Z(\mathrm{d}z)
    =
    \pi_0
    \delta_{z_0}(\mathrm{d}z)
    +
    (1-\pi_0)
    f_{\mathrm{c}}(z)
    \,\mathrm{d}z,
    \label{e:mixed_prior}
\end{align}
where $\delta_{z_0}$ is the Dirac measure at $z_0$.

An observation $Y=y$ obtained under design $\design$ updates both the atom probability and the density within the continuous component.
Define
\begin{align}
    \pi_0(y,\design)
    =
    \mathbb{P}
    \left(
        Z=z_0
        |
        Y=y,\design
    \right),
    \label{e:posterior_atom_probability}
\end{align}
and let $f_{\mathrm{c}}(z|y,\design)$ denote the conditional density of $Z$ given
$Z\neq z_0$, $Y=y$, and $\design$.
The posterior law is
\begin{align}
    \mathbb{P}_{Z|Y,\design}(\mathrm{d}z|y,\design)
    &=
    \pi_0(y,\design)
    \delta_{z_0}(\mathrm{d}z)
    \nonumber\\
    &\quad+
    \left[
        1-\pi_0(y,\design)
    \right]
    f_{\mathrm{c}}(z|y,\design)
    \,\mathrm{d}z.
    \label{e:mixed_posterior}
\end{align}

The KL divergence between the posterior and prior laws separates into information about the component membership and information about the value within the continuous component:
\begin{align}
    &D_{\mathrm{KL}}
    \left(
        \mathbb{P}_{Z|Y,\design}(\cdot|y,\design)
        \,||\,
        \mathbb{P}_Z
    \right)
    \nonumber\\
    &\quad=
    D_{\mathrm{KL}}
    \left(
        \operatorname{Bern}
        \left(
            \pi_0(y,\design)
        \right)
        \,||\,
        \operatorname{Bern}(\pi_0)
    \right)
    \nonumber\\
    &\qquad\quad+
    \left[
        1-\pi_0(y,\design)
    \right]
    D_{\mathrm{KL}}
    \left(
        f_{\mathrm{c}}(\cdot|y,\design)
        \,||\,
        f_{\mathrm{c}}
    \right),
    \label{e:KL_mixed_prior_posterior}
\end{align}
provided that
$f_{\mathrm{c}}(\cdot|y,\design)$ is absolutely continuous with respect to $f_{\mathrm{c}}$.
Here and below, a KL divergence written between densities denotes the divergence between the corresponding continuous probability laws.
The first term measures information gained about whether $Z$ lies at the atom, while the second measures information gained about its value conditional on belonging to the continuous component.
This decomposition motivates a variational approximation that represents the two components separately.

\subsection{A unified score for mixed laws}
\label{ss:dominating_measure}

The mixed probability laws above are sufficient to define their KL divergence abstractly.
The variational objective, however, will need to be evaluated through pointwise log ratios at sampled values of $Z$.
Such evaluation must use a probability mass when $Z=z_0$ and a Lebesgue density when $Z\in\mathcal{Z}_{\mathrm{c}}$.
We therefore introduce a common dominating measure that represents both quantities through a single Radon--Nikodym derivative:
\begin{align}
    \nu
    =
    \delta_{z_0}+\mu.
    \label{e:mixed_dominating_measure}
\end{align}
The measure $\nu$ accounts for the atom through $\delta_{z_0}$ and for the continuous component through $\mu$.

The Radon--Nikodym derivative of the prior law with respect to $\nu$ is
\begin{align}
    p_{Z,\nu}(z)
    =
    \frac{
        \mathrm{d}\mathbb{P}_Z
    }{
        \mathrm{d}\nu
    }(z)
    =
    \begin{cases}
        \pi_0,
        & z=z_0,\\[0.3em]
        (1-\pi_0)f_{\mathrm{c}}(z),
        & z\in\mathcal{Z}_{\mathrm{c}}.
    \end{cases}
    \label{e:mixed_prior_RN}
\end{align}
Similarly, the posterior derivative is
\begin{align}
    p_{Z|Y,\design,\nu}(z|y,\design)
    &=
    \frac{
        \mathrm{d}
        \mathbb{P}_{Z|Y,\design}(\cdot|y,\design)
    }{
        \mathrm{d}\nu
    }(z)
    \nonumber\\
    &=
    \begin{cases}
        \pi_0(y,\design),
        & z=z_0,\\[0.3em]
        \left[
            1-\pi_0(y,\design)
        \right]
        f_{\mathrm{c}}(z|y,\design),
        & z\in\mathcal{Z}_{\mathrm{c}}.
    \end{cases}
    \label{e:mixed_posterior_RN}
\end{align}
Thus, these derivatives return a probability mass when evaluated at $z_0$ and a Lebesgue density when evaluated in $\mathcal{Z}_{\mathrm{c}}$.
We write KL divergences between probability laws and use the Radon--Nikodym derivatives only when pointwise ratios or log scores are required.

\subsection{Variational lower bound}
\label{ss:vOED}

For a fixed design $\design$, let
\begin{align}
    \mathbb{Q}_{Z|Y}(\cdot|y;\lambda)
\end{align}
denote a variational approximation to
$\mathbb{P}_{Z|Y,\design}(\cdot|y,\design)$, where
$\lambda\in\mathbb{R}^{n_\lambda}$ contains the variational parameters.
The design is not included explicitly as an input to the variational model.
Instead, the model is fitted separately for each design, so the fitted parameter values generally depend on $\design$.
The design dependence of the fitted approximation is therefore implicit through $\lambda$, rather than represented through an amortized form such as
$\mathbb{Q}_{Z|Y,\design}(\cdot|y,\design;\lambda)$.

Assume that $\mathbb{Q}_{Z|Y}(\cdot|y;\lambda)$ is dominated by $\nu$, and define
\begin{align}
    q_{Z|Y,\nu}(z|y;\lambda)
    =
    \frac{
        \mathrm{d}
        \mathbb{Q}_{Z|Y}(\cdot|y;\lambda)
    }{
        \mathrm{d}\nu
    }(z).
    \label{e:variational_RN_general}
\end{align}

The exact goal-oriented EIG from \eqref{e:GO_EIG} can be written as
\begin{align}
    U_{\mathrm{GO}}(\design)
    =
    \mathbb{E}_{Z,Y|\design}
    \left[
        \log
        \frac{
            p_{Z|Y,\design,\nu}(Z|Y,\design)
        }{
            p_{Z,\nu}(Z)
        }
    \right].
    \label{e:GO_EIG_RN}
\end{align}
The Barber--Agakov objective replaces the generally intractable posterior derivative
$p_{Z|Y,\design,\nu}$ with the variational derivative
$q_{Z|Y,\nu}$ in the EIG~\cite{Barber2003,Foster2019,Dong2025}:
\begin{align}
    U_L(\design;\lambda)
    =
    \mathbb{E}_{Z,Y|\design}
    \left[
        \log
        \frac{
            q_{Z|Y,\nu}(Z|Y;\lambda)
        }{
            p_{Z,\nu}(Z)
        }
    \right].
    \label{e:GO_UL}
\end{align}
For each design, the fitted variational parameters satisfy
\begin{align}
    \lambda^\ast(\design)
    \in
    \argmax_{\lambda}
    U_L(\design;\lambda).
    \label{e:lambda_design_dependence}
\end{align}
Thus, although $\design$ is not an explicit input to
$q_{Z|Y,\nu}$, the optimized approximation depends on the design through
$\lambda^\ast(\design)$.

Subtracting \eqref{e:GO_UL} from \eqref{e:GO_EIG_RN} gives
\begin{align}
    &U_{\mathrm{GO}}(\design)
    -
    U_L(\design;\lambda)
    \nonumber\\
    &=
    \mathbb{E}_{Z,Y|\design}
    \left[
        \log
        \frac{
            p_{Z|Y,\design,\nu}(Z|Y,\design)
        }{
            q_{Z|Y,\nu}(Z|Y;\lambda)
        }
    \right]
    \nonumber\\
    &=
    \mathbb{E}_{Y|\design}
    \left[
        D_{\mathrm{KL}}
        \left(
            \mathbb{P}_{Z|Y,\design}(\cdot|Y,\design)
            \,||\,
            \mathbb{Q}_{Z|Y}(\cdot|Y;\lambda)
        \right)
    \right]
    \geq
    0.
    \label{e:UL_bound}
\end{align}
Hence, $U_L(\design;\lambda)$ is a lower bound on
$U_{\mathrm{GO}}(\design)$.
Equality holds if and only if
\begin{align}
    \mathbb{Q}_{Z|Y}(\cdot|y;\lambda)
    =
    \mathbb{P}_{Z|Y,\design}(\cdot|y,\design)
\end{align}
for $\mathbb{P}_{Y|\design}$-almost every $y$.

A finite lower bound requires
\begin{align}
    \mathbb{P}_{Z|Y,\design}(\cdot|y,\design)
    \ll
    \mathbb{Q}_{Z|Y}(\cdot|y;\lambda)
    \label{e:variational_domination}
\end{align}
for $\mathbb{P}_{Y|\design}$-almost every $y$.
The variational law must therefore assign positive probability to every event having positive probability under the true posterior law.

\subsection{Failure of a purely continuous variational family}
\label{ss:continuous_failure}

Consider a variational family that is purely continuous with respect to Lebesgue measure $\mu$:
\begin{align}
    \mathbb{Q}_{Z|Y}^{\mathrm{c}}(\mathrm{d}z|y;\lambda)
    =
    q_{Z|Y}^{\mathrm{c}}(z|y;\lambda)
    \,\mathrm{d}z,
    \label{e:continuous_q_short}
\end{align}
where
\begin{align}
    q_{Z|Y}^{\mathrm{c}}(z|y;\lambda)
    =
    \frac{
        \mathrm{d}
        \mathbb{Q}_{Z|Y}^{\mathrm{c}}(\cdot|y;\lambda)
    }{
        \mathrm{d}\mu
    }(z)
\end{align}
is its density with respect to Lebesgue measure.
The superscript $\mathrm{c}$ indicates that the entire variational law is purely continuous.
This notation is distinct from $q_{\mathrm{c}}$, introduced below for the continuous-component density of the mixed variational law.

A purely continuous law assigns zero probability to every singleton. In particular,
\begin{align}
    \mathbb{Q}_{Z|Y}^{\mathrm{c}}
    \left(
        \{z_0\}
        |
        y;\lambda
    \right)
    =
    0.
    \label{e:continuous_q_zero_atom}
\end{align}
Whenever the true posterior has positive atom probability,
\begin{align}
    \mathbb{P}_{Z|Y,\design}
    \left(
        \{z_0\}
        |
        y,\design
    \right)
    =
    \pi_0(y,\design)
    >
    0,
    \label{e:non_domination}
\end{align}
the true posterior law is not absolutely continuous with respect to the purely continuous variational law:
\begin{align}
    \mathbb{P}_{Z|Y,\design}(\cdot|y,\design)
    \not\ll
    \mathbb{Q}_{Z|Y}^{\mathrm{c}}(\cdot|y;\lambda).
\end{align}
Consequently,
\begin{align}
    D_{\mathrm{KL}}
    \left(
        \mathbb{P}_{Z|Y,\design}(\cdot|y,\design)
        \,||\,
        \mathbb{Q}_{Z|Y}^{\mathrm{c}}(\cdot|y;\lambda)
    \right)
    =
    +\infty,
    \label{e:infinite_kl_continuous_q}
\end{align}
because the variational law assigns zero probability to an event having positive posterior probability.
Formally, the contribution from the atom is
\begin{align}
    \pi_0(y,\design)
    \log
    \frac{
        \pi_0(y,\design)
    }{
        0
    }
    =
    +\infty.
\end{align}

This failure is structural and cannot be corrected through improved optimization or increased model capacity.
No purely continuous law can assign positive probability to the singleton $\{z_0\}$.
If \eqref{e:infinite_kl_continuous_q} holds for a set of observations with positive prior predictive probability, the correctly defined Barber--Agakov objective equals $-\infty$.
It therefore remains a lower bound, but only as a trivial and unusable one.

The same mismatch can be seen by expressing
$\mathbb{Q}_{Z|Y}^{\mathrm{c}}$ relative to the mixed dominating measure
$\nu=\delta_{z_0}+\mu$.
Its Radon--Nikodym derivative is
\begin{align}
    q_{Z|Y,\nu}^{\mathrm{c}}(z|y;\lambda)
    =
    \begin{cases}
        0,
        & z=z_0,\\[0.3em]
        q_{Z|Y}^{\mathrm{c}}(z|y;\lambda),
        & z\in\mathcal{Z}_{\mathrm{c}}.
    \end{cases}
    \label{e:continuous_q_mixed_RN}
\end{align}
The zero value at $z_0$ is the correct representation of the purely continuous law under $\nu$, because that law assigns no atom probability.

Additionally, a standard continuous-density implementation may evaluate a sample
$z^{(i)}=z_0$ using the Lebesgue density value
$q_{Z|Y}^{\mathrm{c}}(z_0|y^{(i)};\lambda)$.
This quantity is only the height of the continuous density at the numerical location $z_0$.
It is neither a probability mass nor the correct mixed-measure derivative, which satisfies $q_{Z|Y,\nu}^{\mathrm{c}}
    \left(
        z_0
        |
        y^{(i)};\lambda
    \right)
    =
    0$
according to \eqref{e:continuous_q_mixed_RN}.
Replacing this zero mixed-measure derivative by the Lebesgue density value changes the objective and further obscures the underlying measure mismatch.
The resulting empirical criterion is no longer an estimator of the measure-theoretically valid Barber--Agakov lower bound and may exceed the exact EIG.

A valid finite approximation must therefore include an explicit atom and assign a learned probability mass to $z_0$.

\subsection{Mixed variational family}
\label{ss:mixed_variational_family}

We introduce a variational law with the same discrete--continuous structure as the true posterior:
\begin{align}
    \mathbb{Q}_{Z|Y}(\mathrm{d}z|y;\lambda)
    &=
    \widetilde{\pi}_0(y;\lambda)
    \delta_{z_0}(\mathrm{d}z)
    \nonumber\\
    &\quad+
    \left[
        1-\widetilde{\pi}_0(y;\lambda)
    \right]
    q_{\mathrm{c}}(z|y;\lambda)
    \,\mathrm{d}z,
    \label{e:mixed_approx}
\end{align}
where
$\widetilde{\pi}_0(y;\lambda)\in(0,1)$ approximates the posterior atom probability and
$q_{\mathrm{c}}(\cdot|y;\lambda)$ is a normalized conditional density for the continuous component on
$\mathcal{Z}_{\mathrm{c}}$.
Here, the subscript $\mathrm{c}$ denotes the continuous component of the mixed law, rather than an entirely continuous variational law.

The Radon--Nikodym derivative of
$\mathbb{Q}_{Z|Y}(\cdot|y;\lambda)$
with respect to $\nu$ is
\begin{align}
    q_{Z|Y,\nu}(z|y;\lambda)
    =
    \begin{cases}
        \widetilde{\pi}_0(y;\lambda),
        & z=z_0,\\[0.3em]
        \left[
            1-\widetilde{\pi}_0(y;\lambda)
        \right]
        q_{\mathrm{c}}(z|y;\lambda),
        & z\in\mathcal{Z}_{\mathrm{c}}.
    \end{cases}
    \label{e:mixed_variational_RN}
\end{align}
The variational score at the atom is now the learned probability mass
$\widetilde{\pi}_0(y;\lambda)$, while the score in the continuous component is the product of the complementary component probability and the conditional density.

The conditional KL gap between the true and variational posterior laws separates into an atom-probability term and a continuous-component term:
\begin{align}
    &D_{\mathrm{KL}}
    \left(
        \mathbb{P}_{Z|Y,\design}(\cdot|y,\design)
        \,||\,
        \mathbb{Q}_{Z|Y}(\cdot|y;\lambda)
    \right)
    \nonumber\\
    &\quad=
    D_{\mathrm{KL}}
    \left(
        \operatorname{Bern}
        \left(
            \pi_0(y,\design)
        \right)
        \,||\,
        \operatorname{Bern}
        \left(
            \widetilde{\pi}_0(y;\lambda)
        \right)
    \right)
    \nonumber\\
    &\qquad\quad+
    \left[
        1-\pi_0(y,\design)
    \right]
    D_{\mathrm{KL}}
    \left(
        f_{\mathrm{c}}(\cdot|y,\design)
        \,||\,
        q_{\mathrm{c}}(\cdot|y;\lambda)
    \right).
    \label{e:KL_mixed_variational}
\end{align}
The first term measures the discrepancy in the posterior probability of the atom.
The second measures the discrepancy between the true and variational conditional densities given membership in the continuous component.

The gap is finite if
$\widetilde{\pi}_0(y;\lambda)\in(0,1)$ and
$f_{\mathrm{c}}(\cdot|y,\design)$ is absolutely continuous with respect to
$q_{\mathrm{c}}(\cdot|y;\lambda)$.
The mixed approximation therefore reduces posterior learning to two coupled tasks: estimating the conditional probability of the atom and approximating the conditional distribution within the continuous component.

\subsection{Parameterization of the continuous component}
\label{ss:conditional_nf}

The mixed variational formulation does not require a specific representation of the continuous-component density
$q_{\mathrm{c}}(z|y;\lambda)$.
Depending on its dimension and complexity, possible choices include a conditional Gaussian distribution, a Gaussian mixture model, another parametric density family, a kernel or basis representation, a transport map, or a normalizing flow.
The essential requirements are that the model define a normalized conditional density and that its support contain the support of the true continuous posterior.

In this work, we represent
$q_{\mathrm{c}}(z|y;\lambda)$
using a conditional normalizing flow~\cite{Rezende2015,Papamakarios2021,Kobyzev2021}.
The complete variational model combines this conditional NF with the atom-probability model
$\widetilde{\pi}_0(y;\lambda)$.
We refer to the resulting atom-plus-continuous construction as the mixed NF, although the NF itself represents only the continuous component.

NFs belong to the broader class of measure-transport constructions~\cite{Marzouk2016,Spantini2018}.
Both approaches use an invertible map to push a tractable reference law to a more complex target law.
Transport-map methods often emphasize structured transformations, including monotone triangular maps represented through polynomial or other basis expansions.
NF methods commonly emphasize compositions of transformations that permit efficient sampling and density evaluation, with neural networks used to parameterize the transformations and their dependence on the conditioning observation.

Let $V_0$ denote a reference random variable with tractable density $p_0$.
For a realization $v_0$ and observation $y$, define the sequence
\begin{align}
    v_k
    =
    T_k
    \left(
        v_{k-1};
        y,\lambda_k
    \right),
    \qquad
    k=1,\ldots,K,
    \label{e:nf_layers}
\end{align}
where each $T_k$ is invertible and differentiable, and
$\lambda_k$ contains the trainable parameters associated with the $k$th transformation.
The resulting composition is
\begin{align}
    T
    =
    T_K\circ\cdots\circ T_1,
\end{align}
and a realization from the modeled continuous component is obtained as
\begin{align}
    z
    =
    T(v_0;y,\lambda).
\end{align}
When $\mathcal{Z}_{\mathrm{c}}$ is constrained, an additional invertible support transformation is included so that the image of $T$ lies in
$\mathcal{Z}_{\mathrm{c}}$.

A conditional affine transformation, for example, can be written as
\begin{align}
    v_k
    =
    \exp
    \left[
        s_k(y;\lambda_k)
    \right]
    v_{k-1}
    +
    t_k(y;\lambda_k),
    \label{e:conditional_affine_layer}
\end{align}
where the scale function $s_k$ and shift function $t_k$ are represented by neural networks conditioned on $y$.
The trainable weights and biases of these networks are contained in $\lambda_k$, while
$s_k(y;\lambda_k)$ and
$t_k(y;\lambda_k)$ are the transformation parameters produced for a given observation.

For a scalar QoI, a composition of affine transformations reduces to a single affine transformation and therefore defines only a conditional location--scale family.
Greater expressivity can be obtained using scalar monotone transformations. In the numerical experiments, each conditional affine map is followed by a simple monotone tanh warp,
\begin{align}
    v_k = v_k + c_k(y;\lambda_k)\tanh(v_k),
\quad |c_k(y;\lambda_k)|<1,
\end{align}
where $c_k(y;\lambda_k)$ is a scalar coefficient produced by a neural network conditioned on $y$. The constraint on $c_k$ keeps the tanh warp monotone and invertible.
The change-of-variables expression below applies to either construction.

For a given
$z\in\mathcal{Z}_{\mathrm{c}}$, let
\begin{align}
    v_0
    =
    T^{-1}(z;y,\lambda).
\end{align}
The conditional continuous-component density is then
\begin{align}
    \log
    q_{\mathrm{c}}(z|y;\lambda)
    =
    \log
    p_0(v_0)
    -
    \sum_{k=1}^{K}
    \log
    \left|
        \frac{
            \partial T_k
        }{
            \partial v_{k-1}
        }
        \left(
            v_{k-1};
            y,\lambda_k
        \right)
    \right|.
    \label{e:NF_density}
\end{align}

A separate neural network represents the conditional atom probability, for example through
\begin{align}
    \widetilde{\pi}_0(y;\lambda_\pi)
    =
    \operatorname{sigmoid}
    \left(
        a_\pi(y;\lambda_\pi)
    \right),
    \label{e:atom_probability_network}
\end{align}
where $a_\pi(y;\lambda_\pi)$ denotes the network output before application of the sigmoid transformation.
The full variational parameter vector collects the parameters of the atom-probability network and the conditional flow networks:
\begin{align}
    \lambda
    =
    \left(
        \lambda_\pi,
        \lambda_1,
        \ldots,
        \lambda_K
    \right).
\end{align}

For a sample pair
$(y^{(i)},z^{(i)})$, the mixed-measure log score is
\begin{align}
    &\log
    q_{Z|Y,\nu}
    \left(
        z^{(i)}
        |
        y^{(i)};\lambda
    \right)
    \nonumber\\
    &\quad=
    \begin{cases}
        \log
        \widetilde{\pi}_0
        \left(
            y^{(i)};\lambda
        \right),
        & z^{(i)}=z_0,\\[0.6em]
        \log
        \left[
            1-
            \widetilde{\pi}_0
            \left(
                y^{(i)};\lambda
            \right)
        \right]
        +
        \log
        q_{\mathrm{c}}
        \left(
            z^{(i)}
            |
            y^{(i)};\lambda
        \right),
        & z^{(i)}\in\mathcal{Z}_{\mathrm{c}}.
    \end{cases}
    \label{e:mixed_log_score}
\end{align}
Samples at the atom are therefore scored using the conditional probability mass.
Samples in the continuous component are scored using the complementary component probability together with the conditional NF density.

\begin{figure*}[pos=t]
    \centering
    \includegraphics[width=0.96\textwidth]{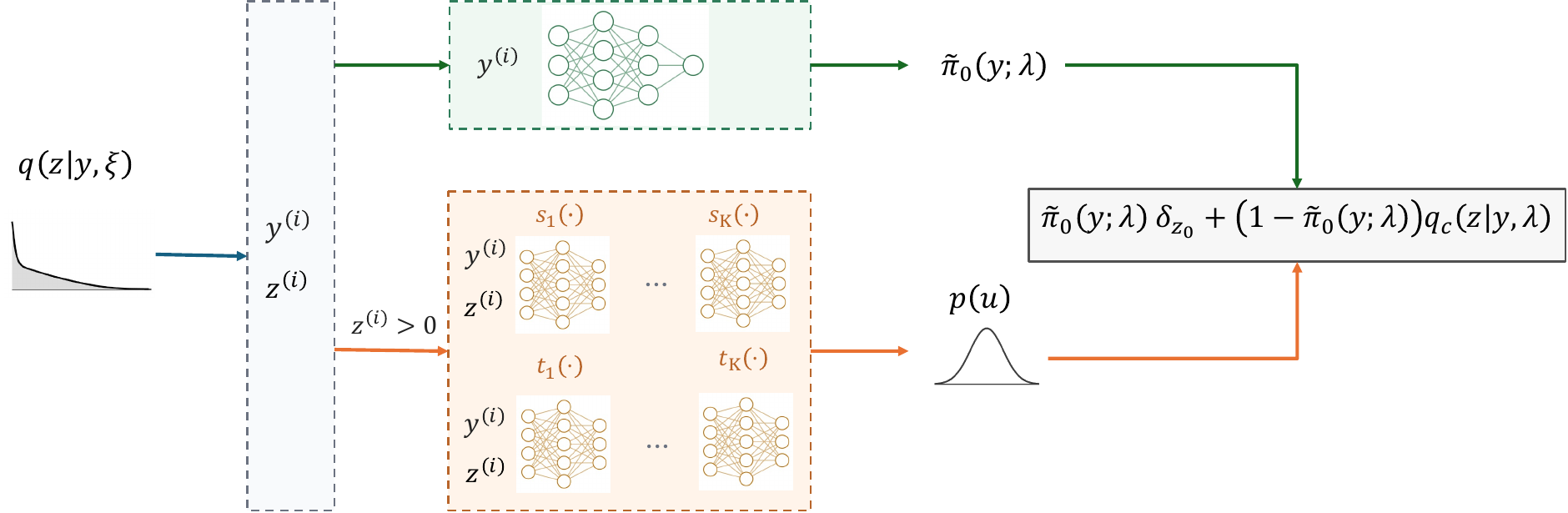}
    \caption{
    Evaluation of the mixed variational posterior law.
    The observation $y^{(i)}$ is provided to neural networks that produce the atom probability
    $\widetilde{\pi}_0(y^{(i)};\lambda)$
    and the parameters of the conditional NF, such as the scale and shift outputs
    $s_k(y^{(i)};\lambda_k)$
    and
    $t_k(y^{(i)};\lambda_k)$.
    A sample satisfying
    $z^{(i)}=z_0$
    is scored using the atom probability.
    A sample in the continuous component is additionally evaluated under the conditional flow density
    $q_{\mathrm{c}}(z^{(i)}|y^{(i)};\lambda)$
    and is scored using
    $\left[1-\widetilde{\pi}_0(y^{(i)};\lambda)\right]
    q_{\mathrm{c}}(z^{(i)}|y^{(i)};\lambda)$.
    }
    \label{fig:nf_diagram}
\end{figure*}

Figure~\ref{fig:nf_diagram} summarizes this evaluation.
The observation is provided to the atom-probability network and to the networks that generate the conditional flow parameters.
A sample at the atom is scored directly using the predicted atom probability.
A sample in the continuous component is additionally passed through the inverse flow and evaluated using the conditional density and the complementary component probability.

\subsection{Monte Carlo estimation and design optimization}
\label{ss:mc_design_optimization}

For a fixed design $\design$, draw joint samples according to
\begin{align}
    \param^{(i)}
    &\sim
    \mathbb{P}_{\Param},
    \nonumber\\
    y^{(i)}
    &\sim
    \mathbb{P}_{Y|\Param,\design}
    \left(
        \cdot
        |
        \param^{(i)},\design
    \right),
    \nonumber\\
    z^{(i)}
    &=
    H
    \left(
        \param^{(i)}
    \right),
    \qquad
    i=1,\ldots,N.
    \label{e:joint_sampling}
\end{align}
The Monte Carlo estimator of the variational objective is
\begin{align}
    \widehat{U}_L(\design;\lambda)
    =
    \frac{1}{N}
    \sum_{i=1}^{N}
    \left[
        \log
        q_{Z|Y,\nu}
        \left(
            z^{(i)}
            |
            y^{(i)};\lambda
        \right)
        -
        \log
        p_{Z,\nu}
        \left(
            z^{(i)}
        \right)
    \right].
    \label{e:Ud_lower_MC}
\end{align}
Samples at $z_0$ contribute log probability masses, whereas samples in
$\mathcal{Z}_{\mathrm{c}}$ contribute log densities with respect to Lebesgue measure.

The overall variational design problem is
\begin{align}
    \design^\ast
    \in
    \argmax_{\design\in\designset}
    \max_{\lambda}
    U_L(\design;\lambda).
    \label{e:UL_OED}
\end{align}
The inner optimization fits the variational approximation at a given design, while the outer optimization compares the resulting optimized lower bounds across designs.
For grid-based optimization, one may train a separate variational model at each candidate design and select the design with the largest optimized value.
For continuous design spaces, the design and variational parameters may instead be updated jointly or alternately, with $\lambda$ adapting to the current value of $\design$.
An explicitly design-conditioned variational model provides an amortized alternative, but is not considered here.

The optimization may be performed using grid search, derivative-free methods, or gradient-based methods.
Grid search is practical for low-dimensional design spaces, while joint or alternating stochastic-gradient ascent becomes more attractive as the dimensions of $\design$ and $\lambda$ increase.

Although stochastic-gradient ascent requires only gradient estimates, the scalar estimator
$\widehat{U}_L(\design;\lambda)$ remains useful.
It can be differentiated automatically to produce stochastic gradients and can also be used to monitor convergence, compare candidate designs, and report the estimated variational objective.

The gradient with respect to the variational parameters is
\begin{align}
    \nabla_{\lambda}
    U_L(\design;\lambda)
    =
    \mathbb{E}_{\Param,Y|\design}
    \left[
        \nabla_{\lambda}
        \log
        q_{Z|Y,\nu}
        \left(
            H(\Param)
            |
            Y;\lambda
        \right)
    \right].
    \label{e:UL_grad_lambda}
\end{align}
The simulated values of $\Param$, $Y$, and $H(\Param)$ are treated as fixed when differentiating with respect to $\lambda$.
This gradient therefore does not require derivatives of either the observation map $G$ or the goal map $H$.

Holding $\lambda$ fixed, one possible design-gradient estimator is obtained using the likelihood-ratio identity:
\begin{align}
    &\quad \nabla_{\design}
    U_L(\design;\lambda) \nonumber \\
    &=
    \mathbb{E}_{\Param,Y|\design}
    \left[
        \log
        q_{Z|Y,\nu}
        \left(
            H(\Param)
            |
            Y;\lambda
        \right)
        \nabla_{\design}
        \log
        p_{Y|\Param,\design}
        \left(
            Y
            |
            \Param,\design
        \right)
    \right],
    \label{e:UL_grad_design}
\end{align}
provided that the regularity conditions for differentiating under the expectation hold.
The prior-law term does not appear because
$\log p_{Z,\nu}(H(\Param))$ is constant with respect to $Y$ conditional on $\Param$, while the conditional expectation of the likelihood score is zero.

The design gradient does not require derivatives of $H$ because $H(\Param)$ has no design dependence in the present formulation.
Evaluation of the likelihood score in \eqref{e:UL_grad_design}, however, requires
$\nabla_{\design}\log p_{Y|\Param,\design}$ to be available.
For an observation model such as
\begin{align}
    Y
    =
    G(\Param,\design)+\Epsilon,
\end{align}
evaluating this score may require $\nabla_{\design}G$.
A pathwise gradient through the same observation model also generally requires $\nabla_{\design}G$.
If these derivatives are unavailable, the design may instead be optimized using finite differences, derivative-free stochastic optimization, Bayesian optimization, or another  method.
Derivatives of $H$ would be required only if the goal map itself depended on $\design$ or if the law of $\Param$ depended on $\design$.

The lower-bound identity in \eqref{e:UL_bound} assumes that the prior law
$\mathbb{P}_Z$, and hence $p_{Z,\nu}$, is known exactly.
If $p_{Z,\nu}$ is replaced by an estimated Radon--Nikodym derivative, the numerical objective contains additional approximation error and may not retain an exact lower-bound guarantee.

Algorithm~\ref{alg:NF_GO_OED} summarizes the complete procedure.

\begin{algorithm}[htbp]
    \caption{Mixed-distribution variational GO-OED}
    \label{alg:NF_GO_OED}
    \begin{algorithmic}[1]
        \REQUIRE
        Prior law $\mathbb{P}_{\Param}$;
        goal map $H$;
        observation law or simulator
        $\mathbb{P}_{Y|\Param,\design}$;
        atom location $z_0$;
        batch size $N$;
        variational models for
        $\widetilde{\pi}_0(y;\lambda)$
        and
        $q_{\mathrm{c}}(z|y;\lambda)$.
        \STATE
        Construct or estimate the prior mixed law $\mathbb{P}_Z$ and its Radon--Nikodym derivative
        $p_{Z,\nu}$ with respect to
        $\nu=\delta_{z_0}+\mu$.
        \STATE
        Initialize the design $\design$ and variational parameters $\lambda$.
        \REPEAT
            \STATE
            Draw
            $\param^{(i)}\sim\mathbb{P}_{\Param}$,
            simulate
            $y^{(i)}\sim
            \mathbb{P}_{Y|\Param,\design}
            (\cdot|\param^{(i)},\design)$,
            and compute
            $z^{(i)}=H(\param^{(i)})$,
            for $i=1,\ldots,N$.
            \STATE
            Evaluate
            $\log q_{Z|Y,\nu}(z^{(i)}|y^{(i)};\lambda)$
            using probability mass when $z^{(i)}=z_0$ and continuous density otherwise.
            \STATE
            Form
            $\widehat{U}_L(\design;\lambda)$
            using \eqref{e:Ud_lower_MC}, or directly estimate the required stochastic gradients.
            \STATE
            Update $\lambda$ and, when applicable, $\design$ using the selected variational and design-optimization procedures.
        \UNTIL{the variational objective and design have stabilized}
        \RETURN
        Optimized design $\design^\ast$ and fitted variational approximation
        $\mathbb{Q}_{Z|Y}(\cdot|y;\lambda^\ast(\design^\ast))$.
    \end{algorithmic}
\end{algorithm}
\section{Analytical Validation Example}
\label{s:toy}

We first consider a controlled example that isolates the mixed-distribution issue from the physical and computational complexities of the ship roll application.
The QoI has the same atom-plus-continuous structure induced by the threshold-based ship safety metric, while the underlying Bayesian inverse problem remains analytically tractable.
The exact goal-oriented EIG therefore provides a reference for assessing the mixed-NF lower bound and illustrating the failure of a purely continuous variational approximation.

\subsection{Problem formulation}
\label{ss:toy_problem}

Let
\begin{align}
    \Param
    =
    \begin{bmatrix}
        \Param_1\\
        \Param_2
    \end{bmatrix}
    \sim
    \mathcal{N}(0,I_2),
\end{align}
and consider the scalar linear observation model
\begin{align}
    Y
    &=
    g(\design)^\top\Param+\Epsilon,
    \qquad
    \Epsilon\sim\mathcal{N}(0,\sigma_\epsilon^2),
    \label{e:toy_observation_model}\\
    g(\design)
    &=
    \begin{bmatrix}
        \cos\design\\
        \sin\design
    \end{bmatrix},
    \qquad
    \design\in\designset=[0,\pi].
    \label{e:toy_design}
\end{align}
The scalar design variable $\design$ determines the direction in parameter space observed by the experiment, while $g(\design)$ is the corresponding observation vector.

We define the QoI as the rectified threshold response
\begin{align}
    Z
    =
    \max\{0,\Param_1-c\},
    \label{e:toy_main_qoi}
\end{align}
where $c$ is a prescribed threshold.
All realizations satisfying $\param_1\leq c$ are mapped to the common value $z_0=0$, whereas realizations satisfying $\param_1>c$ produce continuously varying positive values.
The prior law of $Z$ therefore contains an atom at zero with probability
\begin{align}
    \pi_0
    =
    \mathbb{P}(\Param_1\leq c)
    =
    \Phi(c),
\end{align}
where $\Phi$ denotes the standard Gaussian cumulative distribution function, together with a continuous component on $(0,\infty)$.
A mixed QoI law thus arises even though the prior and observation model are Gaussian and the observation map is linear.

The corresponding goal-oriented EIG is
\begin{align}
    U_{\mathrm{GO}}(\design)
    =
    \mathcal{I}(Z;Y|\design).
    \label{e:toy_GO_utility}
\end{align}
This quantity can be evaluated exactly using the mixed dominating measure
\begin{align}
    \nu
    =
    \delta_0+\mu,
\end{align}
where $\mu$ denotes Lebesgue measure on $(0,\infty)$.
The derivation is provided in Appendix~\ref{app:analytical_mixed_example}.
We use
\begin{align}
    \sigma_\epsilon
    =
    0.35,
    \qquad
    c
    =
    0.25.
\end{align}

The design dependence has a direct interpretation.
At $\design=0$, the observation is a noisy measurement of $\Param_1$, which determines $Z$.
At $\design=\pi$, the observation is a noisy measurement of $-\Param_1$ and therefore contains the same information about $Z$.
These two designs are consequently the most informative.
At $\design=\pi/2$, the observation depends only on $\Param_2$, which is independent of both $\Param_1$ and $Z$.
Hence,
\begin{align}
    U_{\mathrm{GO}}(\pi/2)
    =
    0.
\end{align}

\subsection{Variational approximations}
\label{ss:toy_variational_setup}

For each variational formulation, the design and variational parameters are updated jointly by stochastic gradient ascent using samples generated under the current design. For the variational comparisons, each fixed-design run uses $N=2\times 10^4$ joint samples.

The mixed NF combines a conditional probability mass at $z_0=0$ with a conditional NF for the positive continuous component.
Its sample-wise log score is
\begin{align}
    &\quad \log
    q_{Z|Y,\nu}(z|y;\lambda) \nonumber \\
    &=
    \begin{cases}
        \log \widetilde{\pi}_0(y;\lambda),
        & z=0,\\[0.5em]
        \log
        \left[
            1-\widetilde{\pi}_0(y;\lambda)
        \right]
        +
        \log q_{\mathrm{c}}(z|y;\lambda),
        & z>0.
    \end{cases}
    \label{e:toy_mixed_score}
\end{align}
Thus, outcomes at the atom are scored using probability mass, while outcomes in the continuous component are scored using the complementary component probability and the conditional density.

For comparison, the single-NF formulation models all outcomes with one purely continuous density.
As established in Section~\ref{ss:continuous_failure}, this family assigns zero probability to $\{Z=0\}$, so its measure-theoretically valid Barber--Agakov objective is $-\infty$ whenever the posterior atom probability is positive.
A finite numerical score is obtained only by evaluating zero-valued samples using the Lebesgue density at zero.
The resulting empirical criterion is no longer an estimator of the valid Barber--Agakov lower bound and may exceed the exact EIG.
The comparison therefore illustrates the consequences of ignoring the mixed measure structure rather than comparing two valid variational lower bounds.

\subsection{Results}
\label{ss:toy_case_results}

\begin{figure*}[pos=t]
    \centering
    \subfloat[Training behavior at the informative design $\design=0$.]{
        \includegraphics[width=0.45\linewidth]{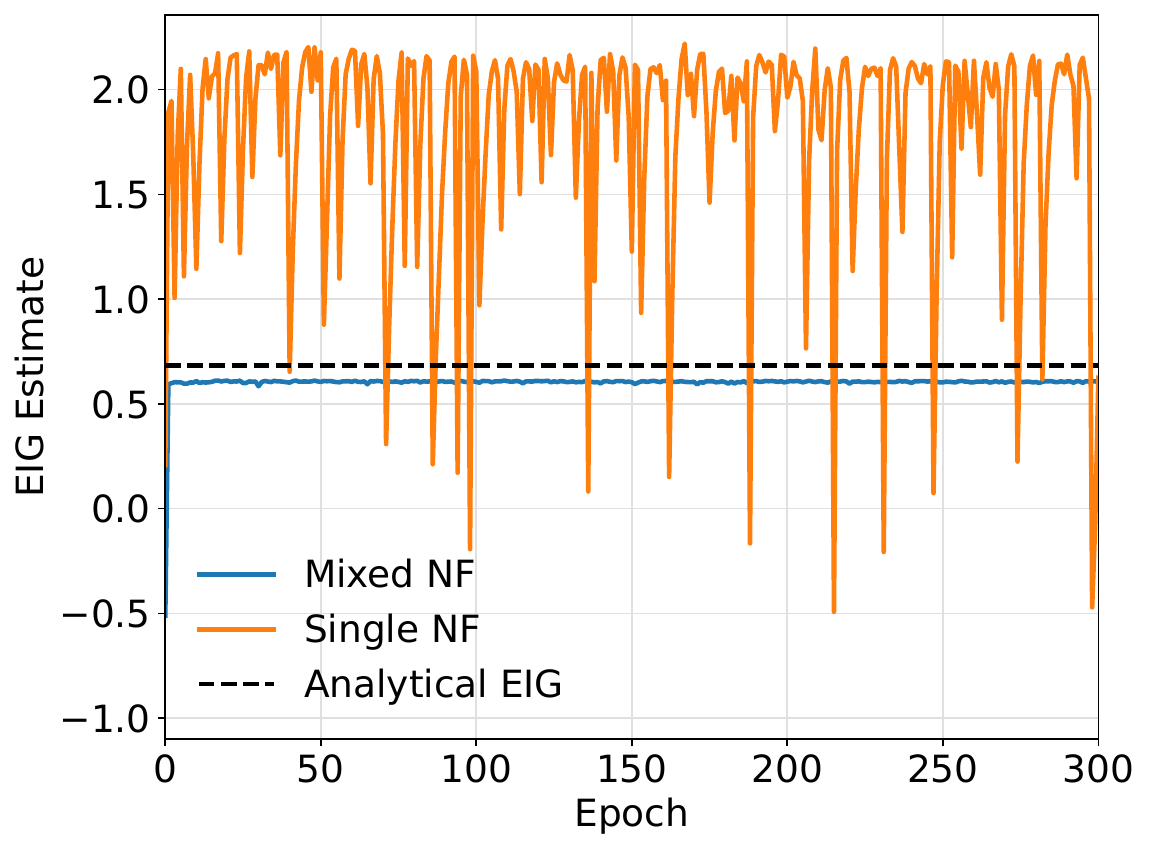}
    }
    \hfill
    \subfloat[Training behavior at the uninformative design $\design=\pi/2$.]{
        \includegraphics[width=0.45\linewidth]{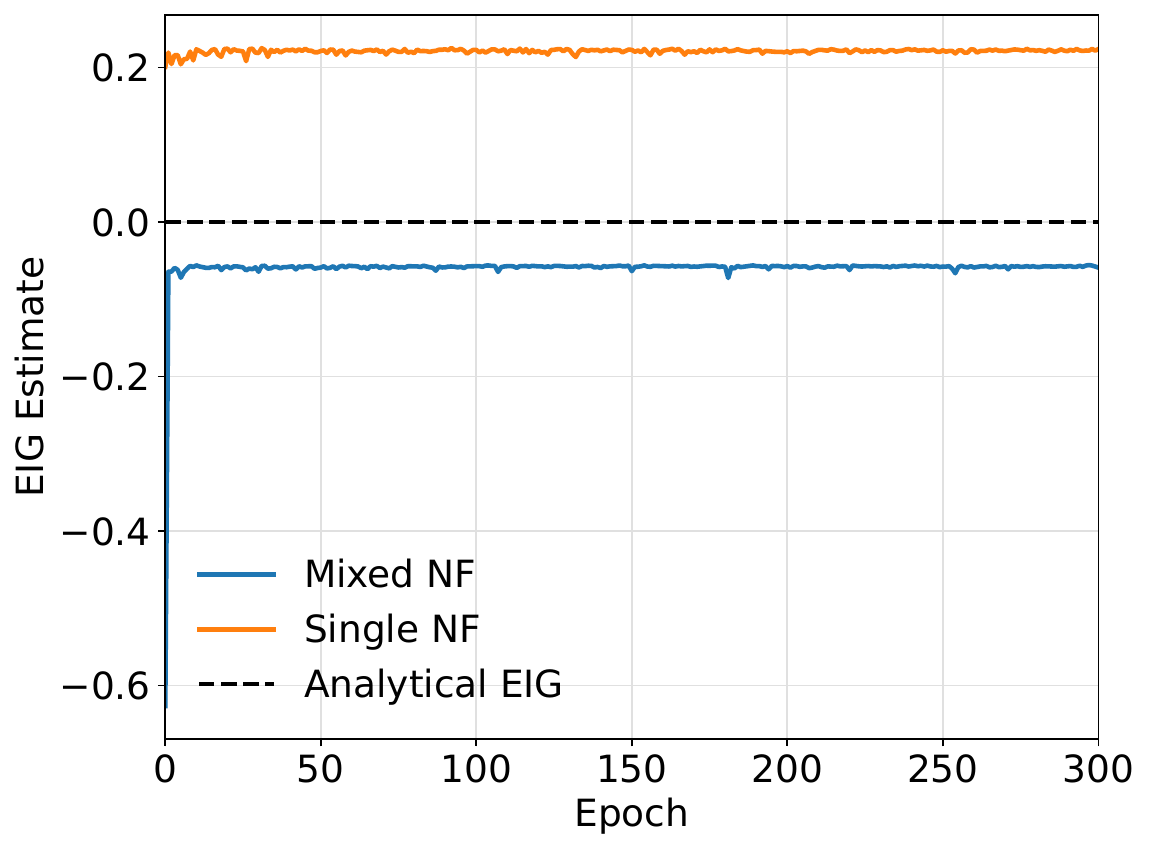}
    }\\[0.9em]
    \subfloat[Analytical goal-oriented EIG, mixed-NF lower-bound estimate, and single-NF score over $\design\in{[0,\pi]}$.]{
        \includegraphics[width=0.7\linewidth]{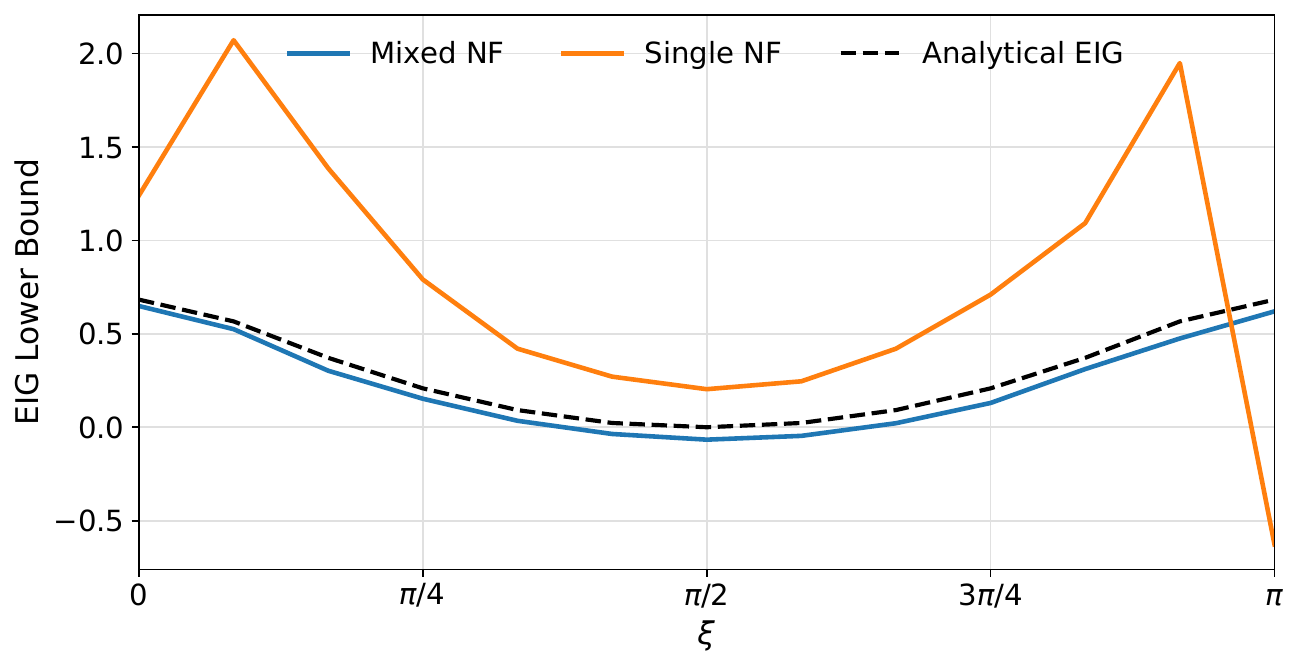}
    }
    \caption{
    Analytical validation of the mixed variational formulation.
    The mixed NF preserves the required atom-plus-continuous structure and captures the correct design dependence.
    The purely continuous NF does not define a valid finite lower bound because it assigns zero probability to the atom at $z=0$.
    }
    \label{fig:toy_variational_results}
\end{figure*}

Figure~\ref{fig:toy_variational_results} compares the analytical goal-oriented EIG with the mixed-NF lower-bound estimate and the single-NF score.
Figures~\ref{fig:toy_variational_results}(a) and~\ref{fig:toy_variational_results}(b) show the training behavior at two diagnostic designs.

At the informative design $\design=0$, the observation is aligned with the parameter component that determines the QoI.
The mixed-NF estimate stabilizes near and below the exact EIG, consistent with the population-level lower-bound identity.
The single-NF score exhibits stronger oscillations and can exceed the analytical EIG.
Such exceedance does not indicate a tighter approximation.
It occurs because the finite single-NF score is not a valid Barber--Agakov lower bound for the mixed posterior law.

At the uninformative design $\design=\pi/2$, the observation is independent of $Z$, and the exact EIG is zero.
The mixed-NF estimate remains just below this reference.
In contrast, the single-NF score remains positive because samples at the atom are evaluated using continuous density values rather than probability masses.
The resulting score can therefore suggest apparent information gain even when $Y$ and $Z$ are independent.

Figure~\ref{fig:toy_variational_results}(c) compares the EIG landscapes over the full design space.
The exact EIG is largest at $\design=0$ and $\design=\pi$, where the observation directly informs $\Param_1$, and vanishes at $\design=\pi/2$, where the observation is independent of the QoI.
The mixed NF reproduces this structure and correctly identifies both the most informative and uninformative designs.
The single-NF score departs from the analytical reference, particularly at designs for which the posterior atom probability varies substantially with the observation.

This example shows that the mixed variational construction is required by the probability law of the QoI itself.
The failure of the purely continuous approximation occurs even in a linear-Gaussian inverse problem with an analytically tractable posterior.
It therefore cannot be attributed to nonlinear dynamics, surrogate-model error, or other complexities of the ship roll application.

\section{Ship Roll Safety Application}
\label{sec:ship}

This section applies the mixed-distribution GO-OED methodology to ship roll safety assessment.
The application combines a nonlinear roll model, a stochastic sea state represented through localized wave groups, and a threshold-based temporal exceedance QoI.
A positive-probability set of ship parameters may produce no threshold exceedance, resulting in an atom at zero in the QoI law.
We first describe the roll dynamics and wave-group representation, then define the safety QoI and wave-group experiment, introduce the surrogate-assisted computational implementation, and present the GO-OED results.

\subsection{Ship dynamics and wave-group representation}
\label{ss:ship_model}

We consider a ship subjected to a stochastic ocean-wave elevation $\eta(t)$ over the finite assessment horizon
$t\in[0,T_{\mathrm{end}}]$.
The roll response $r(t)$ satisfies the nonlinear parameter-dependent dynamical system
\begin{align}
    \ddot{r}
    +
    \mathcal{F}_D(\dot{r};\param)
    +
    \mathcal{F}_R(r;\param)
    =
    \mathcal{M}(r,\eta;\param).
    \label{e:ship_roll_model}
\end{align}
The uncertain parameter vector is
\begin{align}
    \Param
    =
    \begin{bmatrix}
        \Alpha_1 &
        \Alpha_2 &
        \Beta_1 &
        \Beta_2 &
        \Epsilon_1 &
        \Epsilon_2
    \end{bmatrix}^{\top}
    \in
    \paramset
    \subset
    \mathbb{R}^{6},
    \label{e:ship_parameters}
\end{align}
with realization
\begin{align}
    \param
    =
    \begin{bmatrix}
        \alpha_1 &
        \alpha_2 &
        \beta_1 &
        \beta_2 &
        \epsilon_1 &
        \epsilon_2
    \end{bmatrix}^{\top}.
\end{align}
The prior distribution is uniform over the parameter ranges listed in Table~\ref{tab:parameters}.

The damping contribution is
\begin{align}
    \mathcal{F}_D(\dot{r};\param)
    =
    \alpha_1\dot{r}
    +
    \alpha_2\dot{r}|\dot{r}|,
\end{align}
where $\alpha_1$ and $\alpha_2$ govern linear and quadratic damping, respectively.
The restoring contribution is
\begin{align}
    \mathcal{F}_R(r;\param)
    =
    \beta_1r+\beta_2r^3,
\end{align}
where $\beta_1$ governs linear stiffness and the negative coefficient $\beta_2$ produces cubic softening that reduces the effective restoring moment at large roll angles.
The excitation contribution is
\begin{align}
    \mathcal{M}(r,\eta;\param)
    =
    -\epsilon_1\cos(\gamma_0)\eta(t)r
    +
    \epsilon_2\sin(\gamma_0)\eta(t),
\end{align}
where $\epsilon_1$ and $\epsilon_2$ govern parametric and direct excitation, respectively.
The encounter angle $\gamma_0$ is treated as known.

\begin{table}[t]
    \centering
    \caption{Uncertain parameters in the nonlinear ship roll model.}
    \label{tab:parameters}
    \begin{tabular}{lll}
        \hline
        Parameter & Range & Description\\
        \hline
        $\alpha_1$   & $[0.2,0.5]$       & Linear damping\\
        $\alpha_2$   & $[0.02,0.15]$     & Quadratic damping\\
        $\beta_1$    & $[0.02,0.08]$     & Linear stiffness\\
        $\beta_2$    & $[-0.15,-0.05]$   & Cubic stiffness\\
        $\epsilon_1$ & $[0.005,0.02]$    & Parametric excitation\\
        $\epsilon_2$ & $[0.005,0.02]$    & Direct excitation\\
        \hline
    \end{tabular}
\end{table}

The target narrow-band sea state is modeled using the Gaussian spectrum
\begin{align}
    F(\omega)
    =
    \frac{H_s^2}{16}
    \frac{1}{\sqrt{2\pi}\Delta\omega}
    \exp
    \left[
        -\frac{(\omega-\omega_p)^2}{2\Delta\omega^2}
    \right],
    \label{e:spectrum}
\end{align}
where $H_s$ is the significant wave height,
$\omega_p=2\pi/T_p$ is the peak frequency corresponding to the peak period $T_p$, and
$\Delta\omega$ is the spectral bandwidth.
A discrete realization of the wave elevation is
\begin{align}
    \eta(t)
    &=
    \sum_n
    b_n
    \cos
    \left(
        n\delta\omega\,t+\phi_n
    \right),
    \label{e:wave_series}\\
    b_n
    &=
    \sqrt{
        2F(n\delta\omega)\delta\omega
    },
\end{align}
where $\delta\omega$ is the frequency spacing and the phases $\phi_n$ are mutually independent and uniformly distributed on $[0,2\pi)$.
The coefficients $b_n$ are the Fourier amplitudes of the full stochastic wave realization and are distinct from the peak amplitudes used below to characterize individual wave groups.

Narrow-band wave fields naturally organize into localized amplitude-modulated groups.
Following Gong et al.~\cite{Gong2022}, we compute the Hilbert envelope
\begin{align}
    \rho(t)
    =
    \left|
        \eta(t)
        +
        \mathrm{i}\mathcal{H}[\eta(t)]
    \right|,
\end{align}
where $\mathcal{H}[\cdot]$ denotes the Hilbert transform.
Localized features of this envelope are fitted using Gaussian functions of the form
\begin{align}
    \rho_i(t)
    =
    a_i
    \exp
    \left[
        -\frac{(t-t_i)^2}{2l_i^2}
    \right],
    \label{e:group}
\end{align}
where the $i$th extracted group is characterized by its temporal location $t_i$, peak envelope amplitude $a_i$, and temporal half-length $l_i$~\cite{Cousins2016,Gong2022}.
A long stochastic wave realization therefore produces a population
\begin{align}
    \left\{
        (t_i,l_i,a_i)
    \right\}_{i=1}^{m}.
\end{align}
Candidate groups are retained only when the fitted Gaussian envelope adequately represents the corresponding localized feature.
This filtering excludes poorly isolated or irregular features for which $(l_i,a_i)$ does not provide a sufficiently accurate reduced description.

The location $t_i$ identifies where a group occurs in the long wave record, whereas its intrinsic shape is characterized by $(l_i,a_i)$.
After translating the group center to zero, a generic group with half-length $l$ and amplitude $a$ is represented by
\begin{align}
    \eta_{\mathrm{g}}(t;l,a)
    =
    a
    \exp
    \left[
        -\frac{t^2}{2l^2}
    \right]
    \cos
    \left(
        \omega_p t+\phi_{\mathrm{g}}
    \right),
    \label{e:localized_wave_group}
\end{align}
where $\phi_{\mathrm{g}}$ is the prescribed carrier phase.
The amplitude $a$ scales the localized wave elevation, while $l$ controls the temporal width of its envelope.
Substituting \eqref{e:localized_wave_group} into \eqref{e:ship_roll_model} gives
\begin{align}
    \ddot{r}
    +
    \mathcal{F}_D(\dot{r};\param)
    +
    \mathcal{F}_R(r;\param)
    =
    \mathcal{M}
    \left(
        r,
        \eta_{\mathrm{g}}(\cdot;l,a);
        \param
    \right).
    \label{e:group_roll_model}
\end{align}
We denote the resulting response by $r(t;l,a,\param)$ and solve \eqref{e:group_roll_model} over
$t\in[-3l,3l]$ with
\begin{align}
    r(-3l;l,a,\param)
    =
    \dot{r}(-3l;l,a,\param)
    =
    0.
\end{align}

\begin{figure*}[pos=t]
    \centering
    \subfloat[A representative segment of the simulated wave elevation $\eta(t)$, its Hilbert envelope $\rho(t)$, and locally fitted Gaussian wave groups characterized by $(t_i,l_i,a_i)$.]{
        \label{fig:group_parameterization}
        \raisebox{2mm}{
            \includegraphics[width=0.47\linewidth]{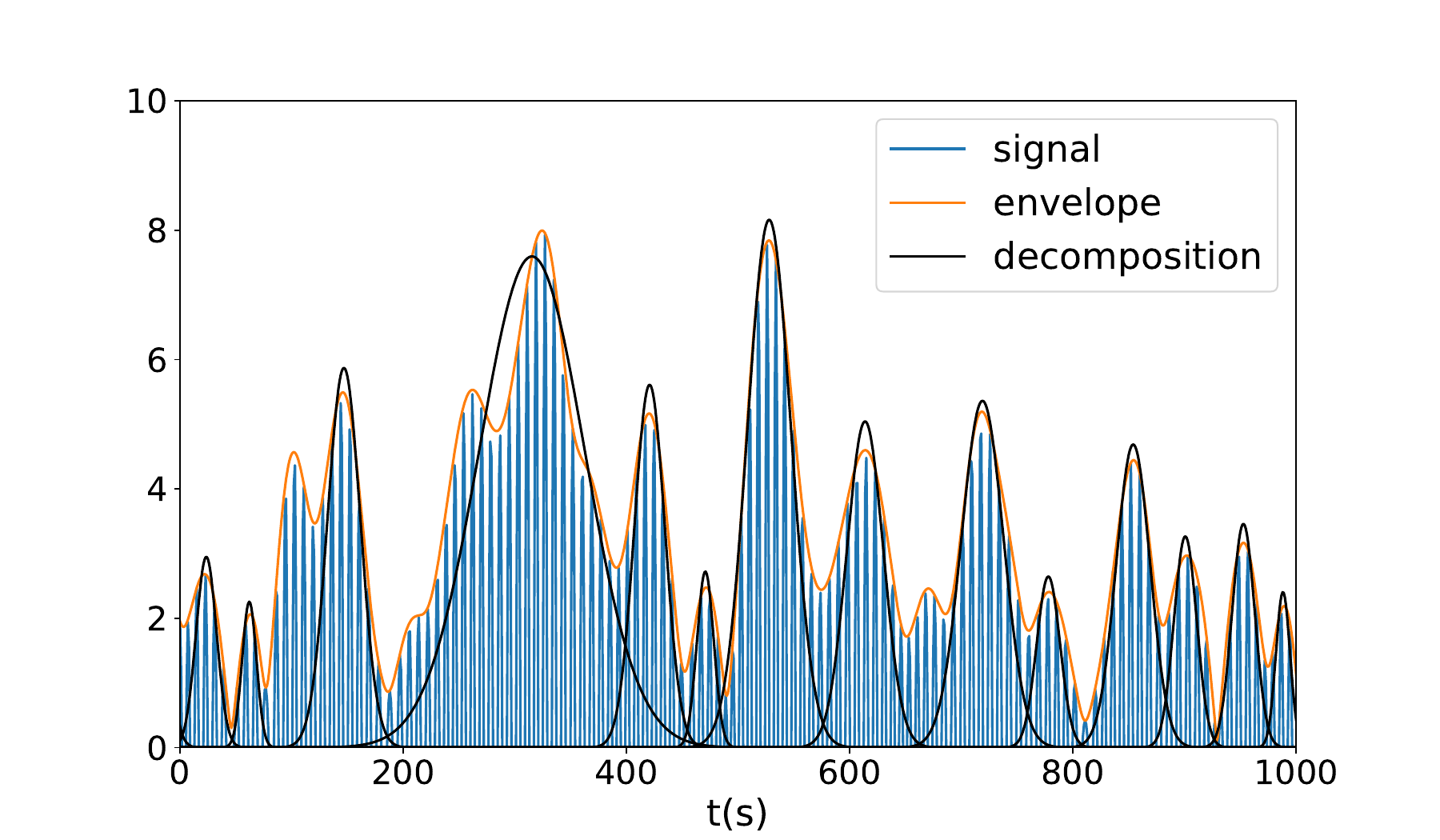}
        }
    }
    \hfill
    \hspace{0.1em}
    \subfloat[Empirical joint density $p_{L,A}(l,a)$ estimated from all retained groups extracted from the full wave realization. The half-length and amplitude axes are normalized by $T_p$ and $H_s$, respectively.]{
        \label{fig:group_density}
        \includegraphics[width=0.47\linewidth]{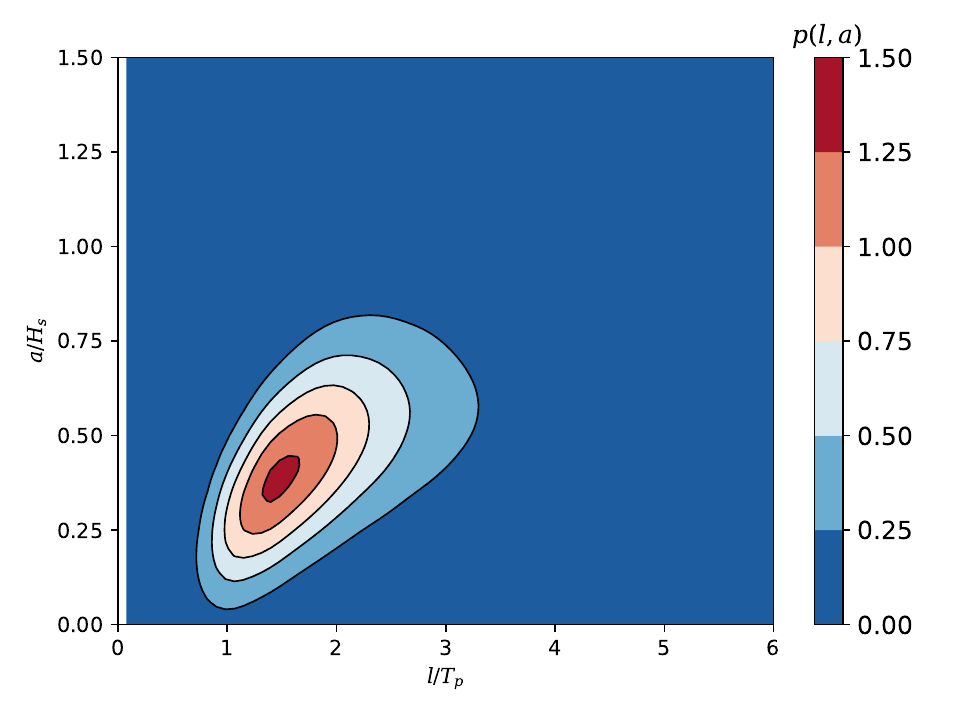}
    }
    \caption{
        Wave-group representation of the target sea state.
    }
    \label{fig:wave}
\end{figure*}

The wave-group representation reduces the stochastic sea state to a population of localized events characterized by the random pair $(L,A)$.
The joint density $p_{L,A}(l,a)$ is estimated from all retained groups, as illustrated in Figure~\ref{fig:wave}.
Lowercase $(l,a)$ denotes a generic realization from this distribution, while $(l_i,a_i)$ denotes the parameters of the individual extracted groups.

\subsection{Safety QoI and wave-group experimental design}
\label{ss:ship_qoi_design}

The safety QoI is the temporal exceedance probability
\begin{align}
    P_{\mathrm{temp}}(\param)
    =
    \frac{1}{T_{\mathrm{end}}}
    \int_0^{T_{\mathrm{end}}}
    \mathbf{1}
    \left(
        |r(t;\param)|>r_s
    \right)
    \,\mathrm{d}t,
    \label{e:Ptemp}
\end{align}
where $r_s$ is the prescribed roll-angle threshold and $\mathbf{1}(\cdot)$ denotes the indicator function.
This quantity is the fraction of the assessment horizon during which the absolute roll angle exceeds the safety threshold.

Direct evaluation of \eqref{e:Ptemp} requires a long time-domain simulation under the stochastic sea state.
The wave-group representation provides a reduced approximation~\cite{Gong2022}.
For a localized group with half-length $l$ and amplitude $a$, define the exceedance duration
\begin{align}
    S(l,a;\param)
    =
    \int_{-3l}^{3l}
    \mathbf{1}
    \left(
        |r(t;l,a,\param)|>r_s
    \right)
    \,\mathrm{d}t.
    \label{e:group_exceedance_duration}
\end{align}
The contribution of the extracted wave groups to the temporal exceedance probability is approximated by
\begin{align}
    z=H(\param)
    \coloneqq
    P_{\mathrm{temp}}^{\mathrm{a}}(\param)
    &=
    \frac{1}{T_{\mathrm{end}}}
    \sum_{i=1}^{m}
    S(l_i,a_i;\param)
    \nonumber\\
    &\approx
    \frac{m}{T_{\mathrm{end}}}
    \iint
    S(l,a;\param)
    p_{L,A}(l,a)
    \,\mathrm{d}l\,\mathrm{d}a.
    \label{e:z_integral}
\end{align}
We use $P_{\mathrm{temp}}^{\mathrm{a}}$ as an approximation of $P_{\mathrm{temp}}$ throughout the application.
Its accuracy depends on the wave-group representation, the localized carrier-wave model in \eqref{e:localized_wave_group}, and the empirical approximation of $p_{L,A}$.

The uncertain parameter $\Param$ induces the random QoI
\begin{align}
    Z
    =
    H(\Param).
\end{align}
The physical roll-angle threshold is $r_s$, whereas the atom in the QoI law occurs at the non-exceedance value $z_0=0$.
Define
\begin{align}
    \paramset_0
    =
    \left\{
        \param\in\paramset:
        S(l,a;\param)=0
        \text{ for }p_{L,A}\text{-almost every }(l,a)
    \right\}.
    \label{e:ship_safe_set}
\end{align}
Every $\param\in\paramset_0$ satisfies $z(\param)=0$.
Consequently, if
\begin{align}
    \mathbb{P}_{\Param}(\paramset_0)>0,
\end{align}
then the prior law of $Z$ contains an atom at zero.
Under the assumed model, parameter realizations outside $\paramset_0$ produce positive QoI values that form a continuous component.
The resulting prior law has the mixed form
\begin{align}
    \mathbb{P}_Z(\mathrm{d}z)
    =
    \pi_0\delta_0(\mathrm{d}z)
    +
    (1-\pi_0)
    f_{\mathrm{c}}(z)
    \,\mathrm{d}z.
    \label{e:ship_mixed_prior}
\end{align}

A candidate experiment applies a single localized wave group characterized by
\begin{align}
    \design
    =
    (l,a)
    \in
    \designset,
    \label{e:ship_design}
\end{align}
where $l$ and $a$ are the controllable half-length and peak amplitude.
The target sea state contains a population of groups, but each experiment uses one selected pair $(l,a)$.
The design space is restricted to the effective support of $p_{L,A}$ so that candidate experiments correspond to wave groups represented in the target sea state.

For a candidate experiment with wave-group design $\design=(l,a)$, the full roll trajectory $r(t;l,a,\param)$ could serve as the observation, but it is time dependent.
The scalar exceedance duration $S(l,a;\param)$ is more compact, but it is identically zero throughout the non-exceedance regime.
Using $S$ directly would therefore assign the same response to parameter realizations whose maximum roll angles may lie at substantially different distances below the threshold.
Such differences remain informative about whether the safety threshold is likely to be exceeded under the broader wave-group population.

We therefore define the scalar observation map
\begin{align}
    G(\param,\design)
    =
    \begin{cases}
        S(l,a;\param)/l,
        & S(l,a;\param)>0,\\[0.4em]
        \bigl[r_{\max}(l,a;\param)-r_s\bigr]/r_s,
        & S(l,a;\param)=0,
    \end{cases}
    \label{e:G_def}
\end{align}
where $\design=(l,a)$ and
\begin{align}
    r_{\max}(l,a;\param)
    =
    \max_{t\in[-3l,3l]}
    |r(t;l,a,\param)|.
\end{align}
In the exceedance regime, $G$ records the exceedance duration normalized by the group half-length.
In the non-exceedance regime, it records the normalized margin between the maximum roll response and the safety threshold.
The map therefore remains informative on both sides of the threshold while reducing the experimental response to a scalar.

The exceedance duration is recovered from $G$ through
\begin{align}
    S(l,a;\param)
    =
    l\max\left\{0,G(\param,\design)\right\}.
    \label{e:S_from_G}
\end{align}
The auxiliary observation map therefore preserves the quantity required to evaluate the temporal exceedance QoI.

The noisy observation model is
\begin{align}
    Y
    =
    G(\Param,\design)
    +
    \Epsilon,
    \qquad
    \Epsilon
    \sim
    \mathcal{N}(0,\sigma_\epsilon^2),
    \label{e:obs_model}
\end{align}
with likelihood
\begin{align}
    p_{Y|\Param,\design}(y|\param,\design)
    =
    \mathcal{N}
    \left(
        y;
        G(\param,\design),
        \sigma_\epsilon^2
    \right).
    \label{e:ship_likelihood}
\end{align}
The GO-OED objective is
\begin{align}
    U_{\mathrm{GO}}(\design)
    =
    \mathcal{I}(Z;Y|\design),
\end{align}
and the selected wave group satisfies
\begin{align}
    \design^\ast
    =
    (l^\ast,a^\ast)
    \in
    \argmax_{\design\in\designset}
    U_{\mathrm{GO}}(\design).
    \label{e:ship_optimal_design}
\end{align}
The resulting experiment is selected to produce an observation that is informative about the temporal exceedance QoI rather than the full parameter vector.

\subsection{Surrogate-assisted computational implementation}
\label{ss:ship_implementation}

We fix
\begin{align}
    H_s &= 12\,\mathrm{m},
    &
    T_p &= 15\,\mathrm{s},
    &
    \Delta\omega &= 0.02\,\mathrm{s}^{-1},
    \nonumber\\
    \delta\omega &= 0.00026\,\mathrm{s}^{-1},
    &
    \gamma_0 &= \pi/6.
\end{align}
All six roll-model parameters in \eqref{e:ship_parameters} are treated as uncertain. The observation-noise standard deviation is set to $\sigma_\epsilon = 0.05$.

\begin{figure}[pos=t]
    \centering
    \includegraphics[width=\linewidth]{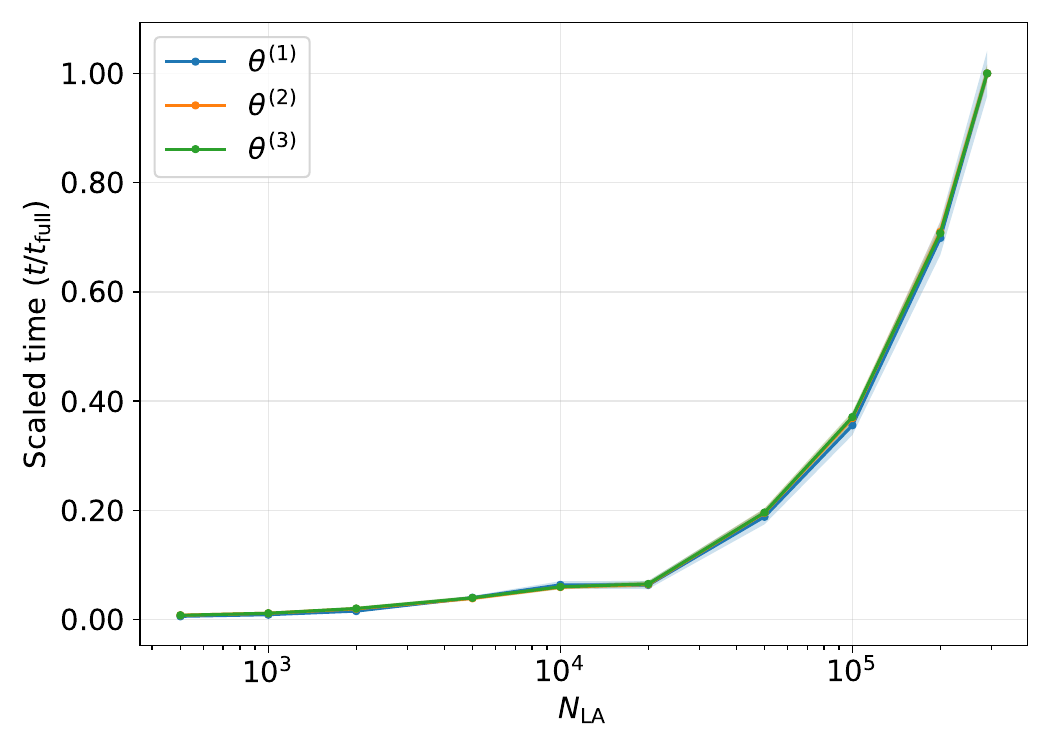}\\
    \includegraphics[width=\linewidth]{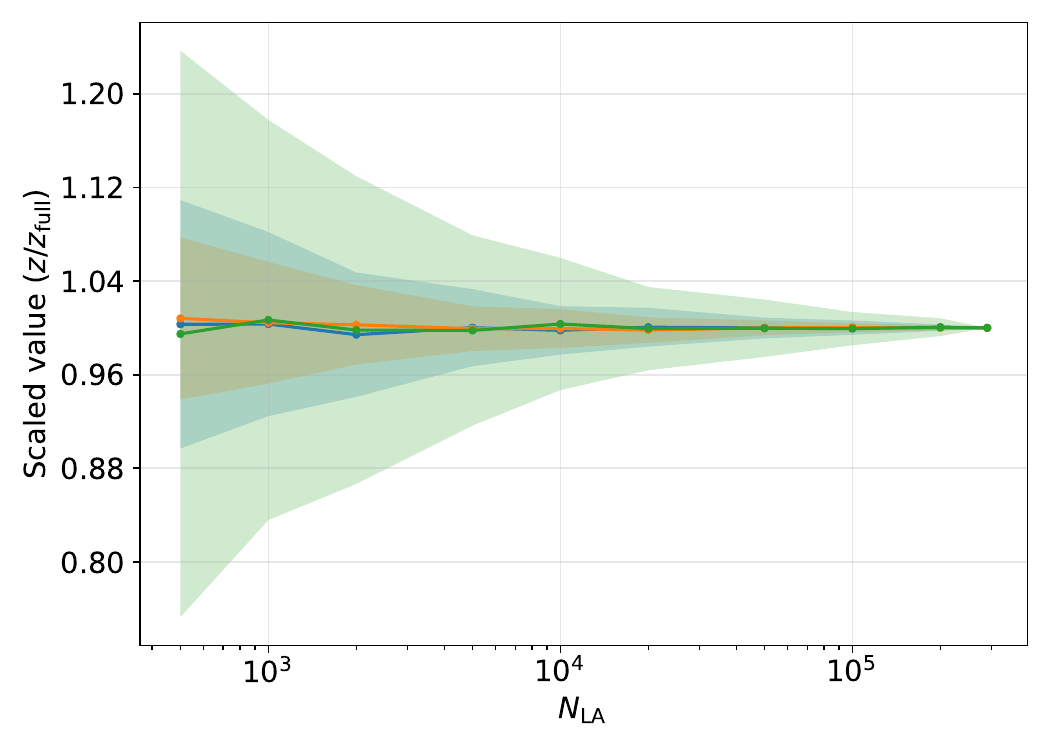}
    \caption{
        Effect of the common wave-group sample size $N_{\mathrm{LA}}$ on evaluation of the safety QoI.
        The upper panel reports scaled wall-clock time for three representative parameter realizations.
        The lower panel reports scaled $z$, with the shaded region indicating variation across 100 independently generated sample sets.
    }
    \label{fig:LA_samples}
\end{figure}

Repeated evaluation of \eqref{e:z_integral} is required for many parameter realizations.
We approximate the wave-group integral using the fixed common sample set
\begin{align}
    \mathcal{W}
    =
    \left\{
        (l_j,a_j)
    \right\}_{j=1}^{N_{\mathrm{LA}}},
    \qquad
    (l_j,a_j)
    \sim
    p_{L,A}.
    \label{e:common_wave_group_sample}
\end{align}
For a parameter realization $\param^{(i)}$, the resulting approximation is
\begin{align}
    z^{(i)}
    =
    \frac{m}{
        N_{\mathrm{LA}}T_{\mathrm{end}}
    }
    \sum_{j=1}^{N_{\mathrm{LA}}}
    l_j
    \max
    \left\{
        0,
        G\bigl(\param^{(i)},(l_j,a_j)\bigr)
    \right\}.
    \label{e:z_CRN}
\end{align}
The experimental design consists of one controllable wave group, whereas the QoI remains an average over the full wave-group population represented by $\mathcal{W}$.
Reusing the same sample set for all parameter realizations reduces variation associated with repeatedly discretizing the sea-state distribution.
Figure~\ref{fig:LA_samples} examines the computational cost and convergence of this approximation.
We use $N_{\mathrm{LA}}=10^4$ in the subsequent analyses.

Direct evaluation of $G(\param,\design)$ requires solving the nonlinear roll equation for each parameter--wave-group pair, which is computationally prohibitive at the scale required for variational GO-OED.
We therefore construct an offline, threshold-specific surrogate
\begin{align}
    \widehat{G}_{r_s}(\param,\design)
\end{align}
over $\paramset\times\designset$.
Training inputs are sampled from the joint distribution
\begin{align}
    p_{\Param}(\param)p_{L,A}(l,a),
\end{align}
and the corresponding targets are generated using the high-fidelity roll simulator.

Separate surrogates are trained for
\begin{align}
    r_s
    =
    0.25\,\mathrm{rad}
    \qquad\text{and}\qquad
    r_s
    =
    0.1\,\mathrm{rad},
\end{align}
because both the exceedance duration and the non-exceedance margin depend explicitly on the threshold.
Once trained, $\widehat{G}_{r_s}$ is used to evaluate the observation model during variational training.
The same surrogate also enables efficient evaluation of the QoI:
\begin{align}
    \widehat{z}_{r_s}^{(i)}
    =
    \frac{m}{
        N_{\mathrm{LA}}T_{\mathrm{end}}
    }
    \sum_{j=1}^{N_{\mathrm{LA}}}
    l_j
    \max
    \left\{
        0,
        \widehat{G}_{r_s}
        \bigl(\param^{(i)},(l_j,a_j)\bigr)
    \right\}.
    \label{e:hat_z}
\end{align}

\begin{figure*}[pos=t]
    \centering
    \includegraphics[width=\linewidth]{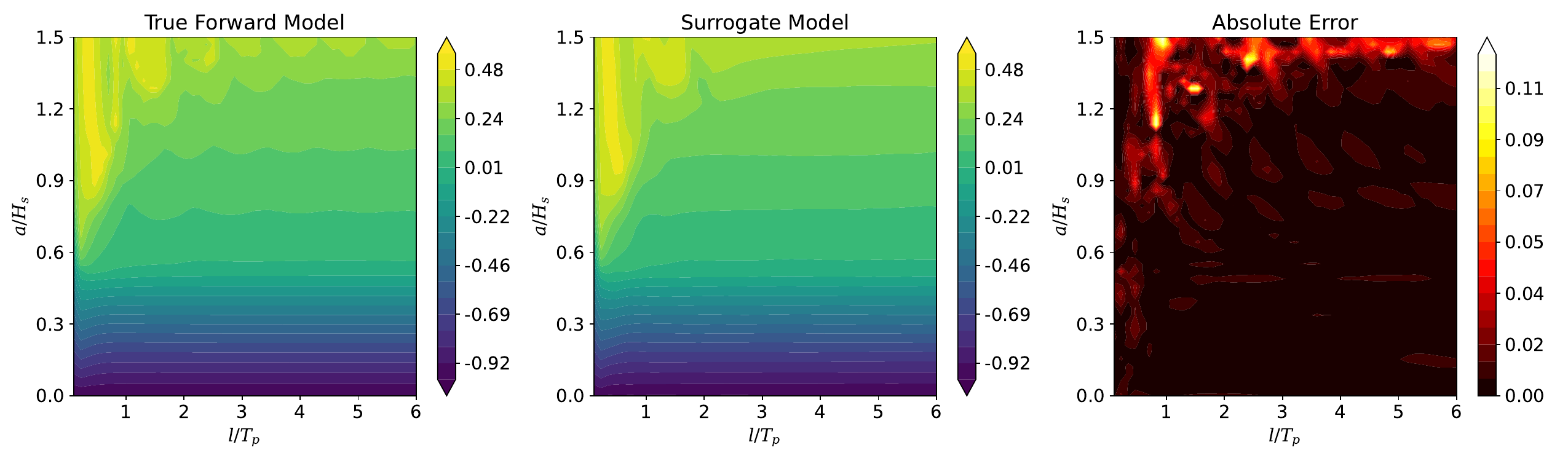}\\
    \includegraphics[width=\linewidth]{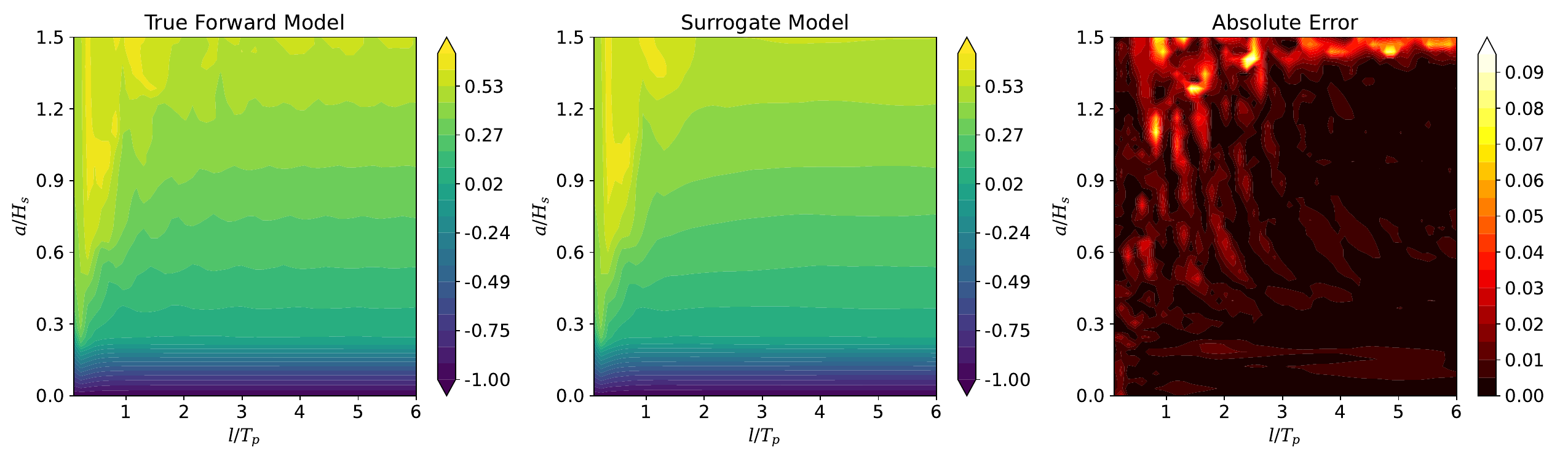}
    \caption{
        Accuracy of the threshold-specific surrogate models for a representative parameter realization.
        The upper row corresponds to $r_s=0.25\,\mathrm{rad}$ and the lower row to $r_s=0.1\,\mathrm{rad}$.
        Each row shows the high-fidelity response, surrogate prediction, and absolute error over the normalized wave-group coordinates $(l/T_p,a/H_s)$.
    }
    \label{fig:surrogate}
\end{figure*}

Figure~\ref{fig:surrogate} compares surrogate predictions with high-fidelity auxiliary responses for a representative parameter realization.
For both thresholds, the surrogate captures the dominant response structure over the wave-group design space, including broad low-response regions and localized high-response regions.
The largest discrepancies occur for short, high-amplitude groups, where the dynamics approach a strongly nonlinear regime and the response surface develops sharper local features~\cite{Gong2022}.
These groups occupy a relatively small portion of the empirical wave-group distribution, limiting their contribution to the integrated QoI in the present study.
The localized discrepancies nevertheless suggest that adaptive sampling in the strongly nonlinear region could improve future surrogate construction.

\begin{figure*}[pos=t]
    \centering
    \subfloat[$r_s=0.25\,\mathrm{rad}$.]{
        \includegraphics[width=0.48\linewidth]{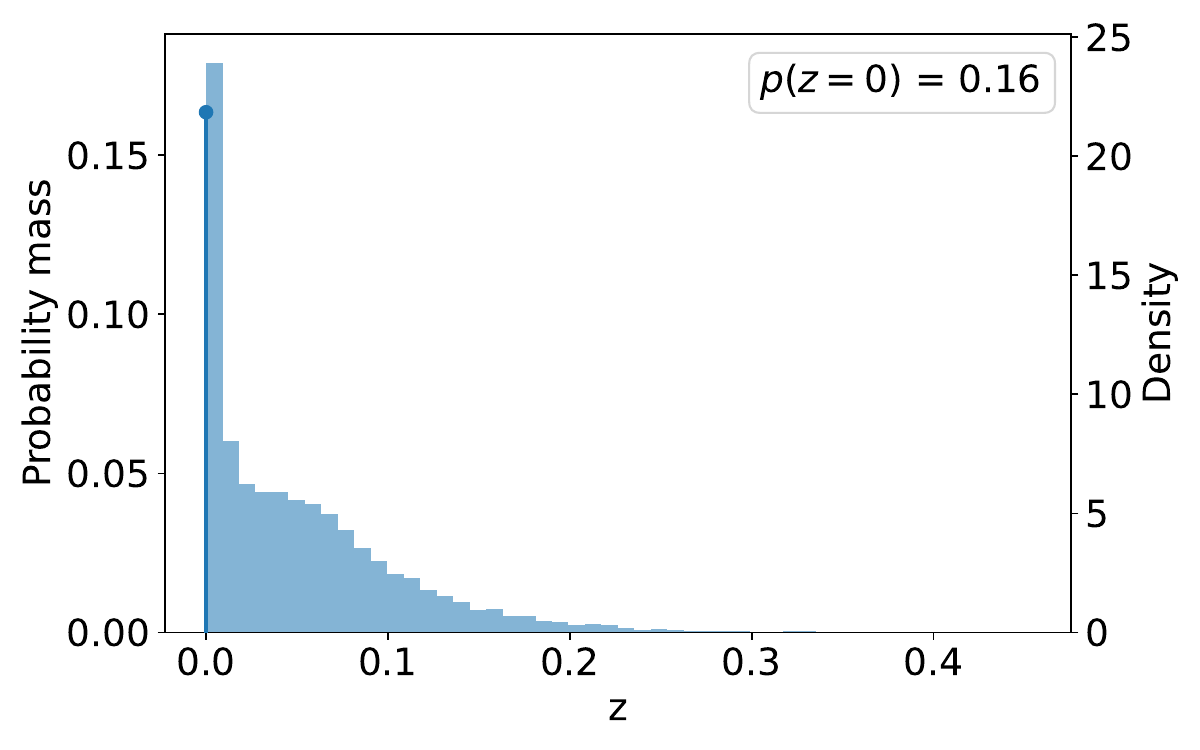}
    }
    \hfill
    \subfloat[$r_s=0.1\,\mathrm{rad}$.]{
        \includegraphics[width=0.48\linewidth]{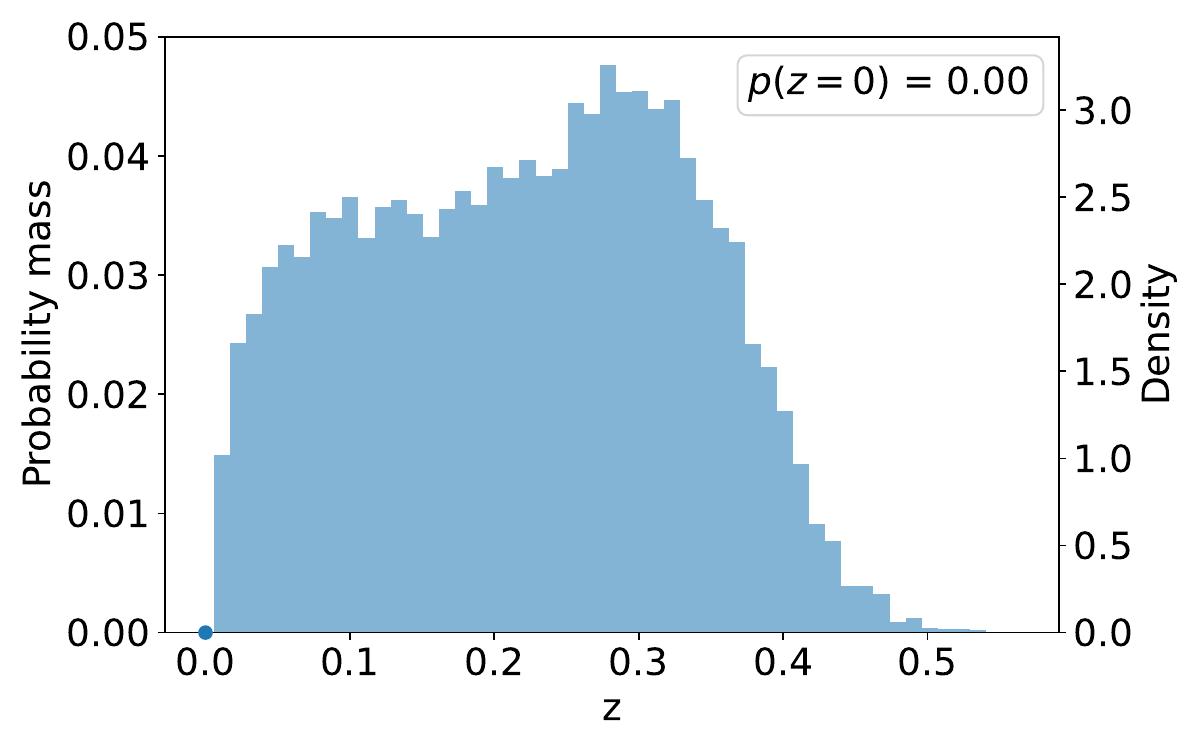}
    }\\
    \caption{
        Prior laws of the ship safety QoI under two roll-angle thresholds, computed using the same prior parameter ensemble.
        For $r_s=0.25\,\mathrm{rad}$, the value at zero represents an empirical probability mass and the positive values form a continuous component.
        For $r_s=0.1\,\mathrm{rad}$, no atom is observed in the sampled ensemble and the law is treated as effectively continuous.
    }
    \label{fig:threshold_distribution_comparison}
\end{figure*}

Figure~\ref{fig:threshold_distribution_comparison} compares the prior laws of $Z$ under the two thresholds using the same prior parameter ensemble.
For $r_s=0.25\,\mathrm{rad}$, a non-negligible subset of sampled parameter realizations produces no exceedance over the common wave-group sample, yielding an empirical atom at zero together with a positive continuous component.
For $r_s=0.1\,\mathrm{rad}$, every sampled parameter realization produces a positive temporal exceedance probability.
No atom is observed in the sampled prior ensemble, and the resulting law is treated as effectively continuous.
This comparison changes the threshold-dependent QoI and observation map while retaining the same roll model, prior distribution, wave-group population, design space, and computational procedure.

\subsection{GO-OED results}
\label{ss:ship_results}

\paragraph{Variational training behavior.}

\begin{figure*}[pos=t]
    \centering
    \subfloat[$r_s=0.25\,\mathrm{rad}$, $l=1.5$, $a=8.0$.]{
        \includegraphics[width=0.45\linewidth]{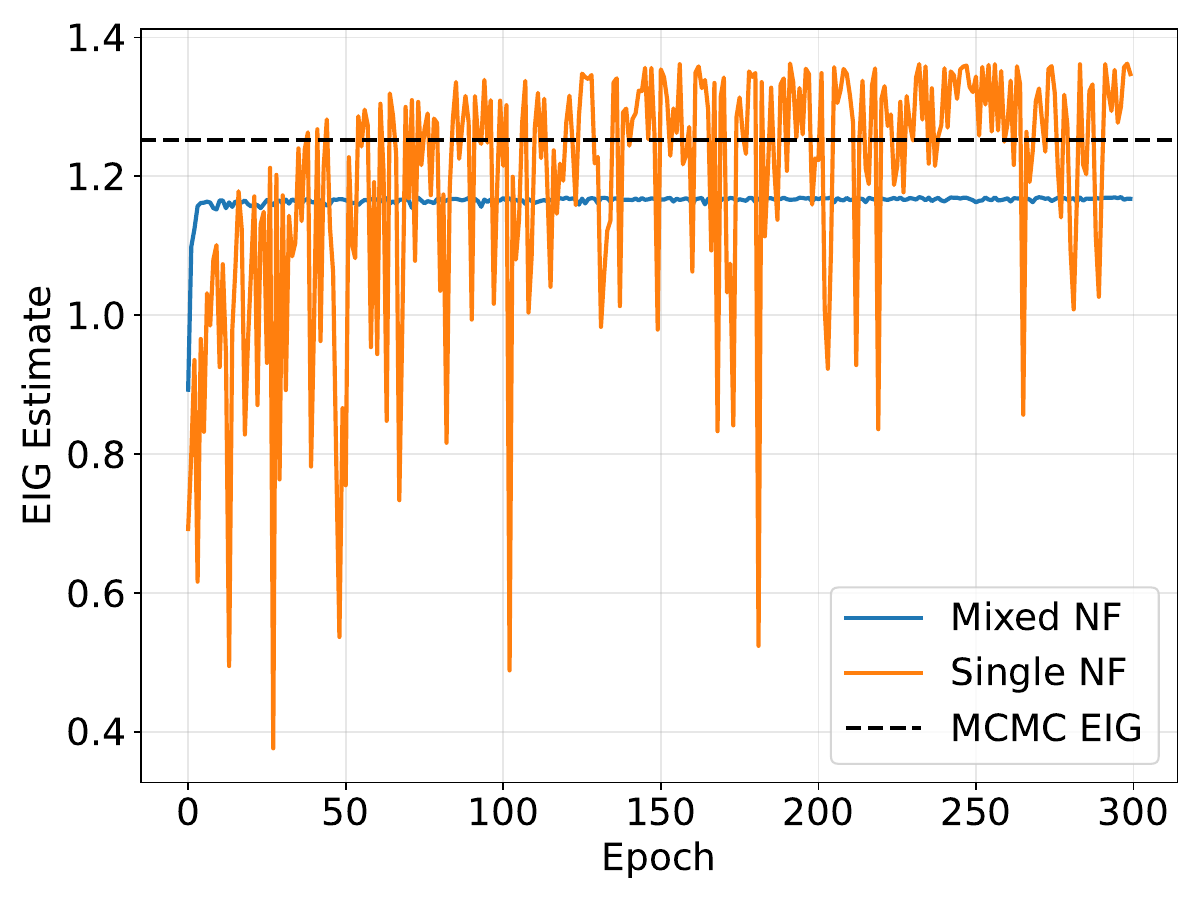}
    }
    \hfill
    \subfloat[$r_s=0.25\,\mathrm{rad}$, $l=40.8$, $a=0$.]{
        \includegraphics[width=0.45\linewidth]{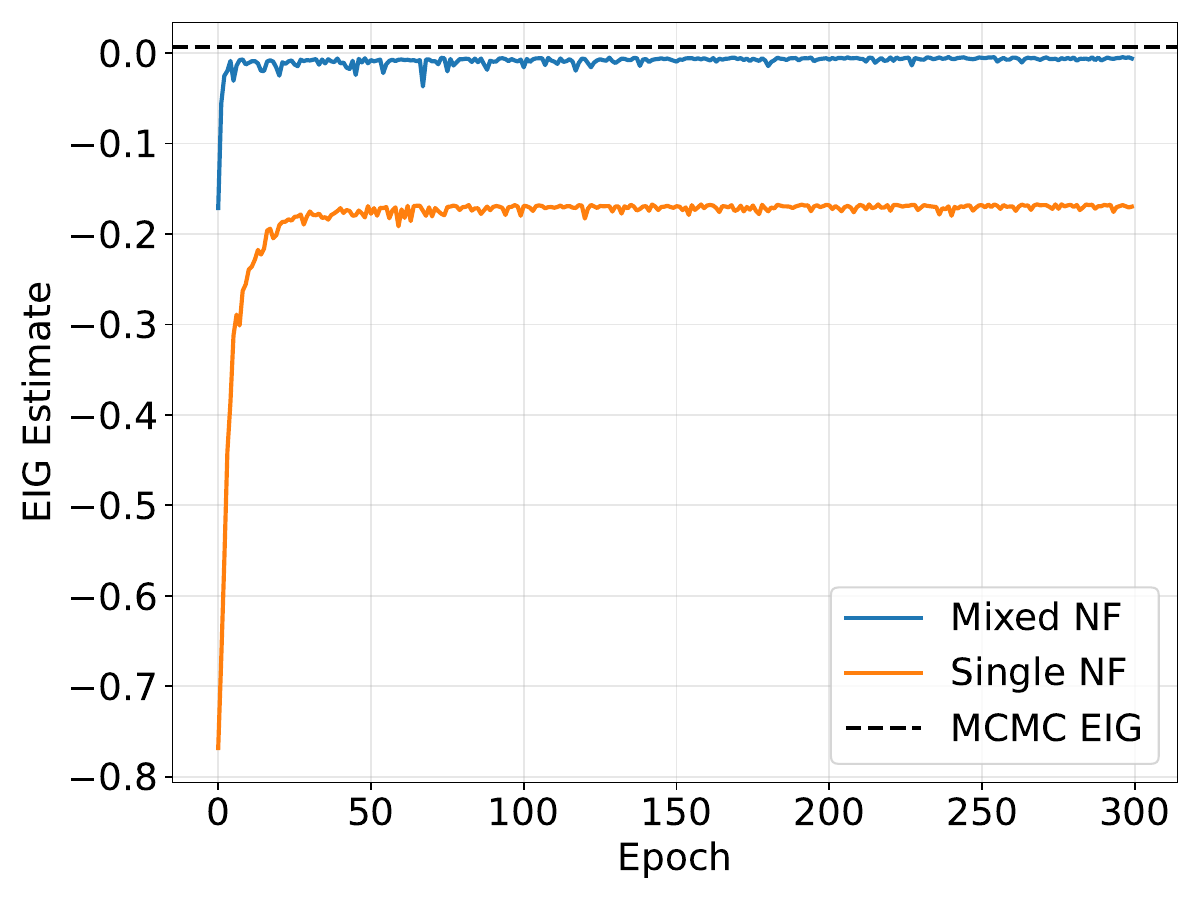}
    }\\
    \subfloat[$r_s=0.1\,\mathrm{rad}$, $l=1.5$, $a=0$.]{
        \includegraphics[width=0.45\linewidth]{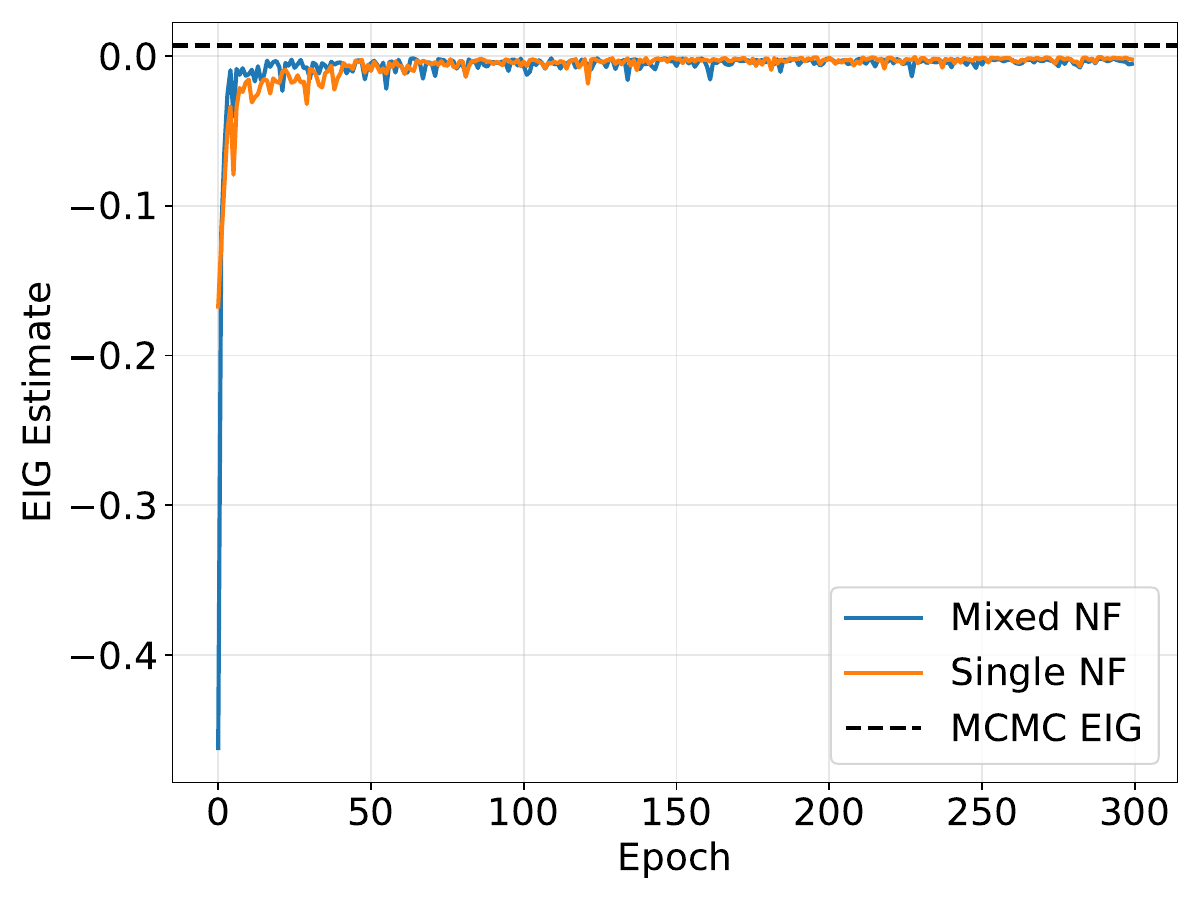}
    }
    \hfill
    \subfloat[$r_s=0.1\,\mathrm{rad}$, $l=50.7$, $a=10.0$.]{
        \includegraphics[width=0.45\linewidth]{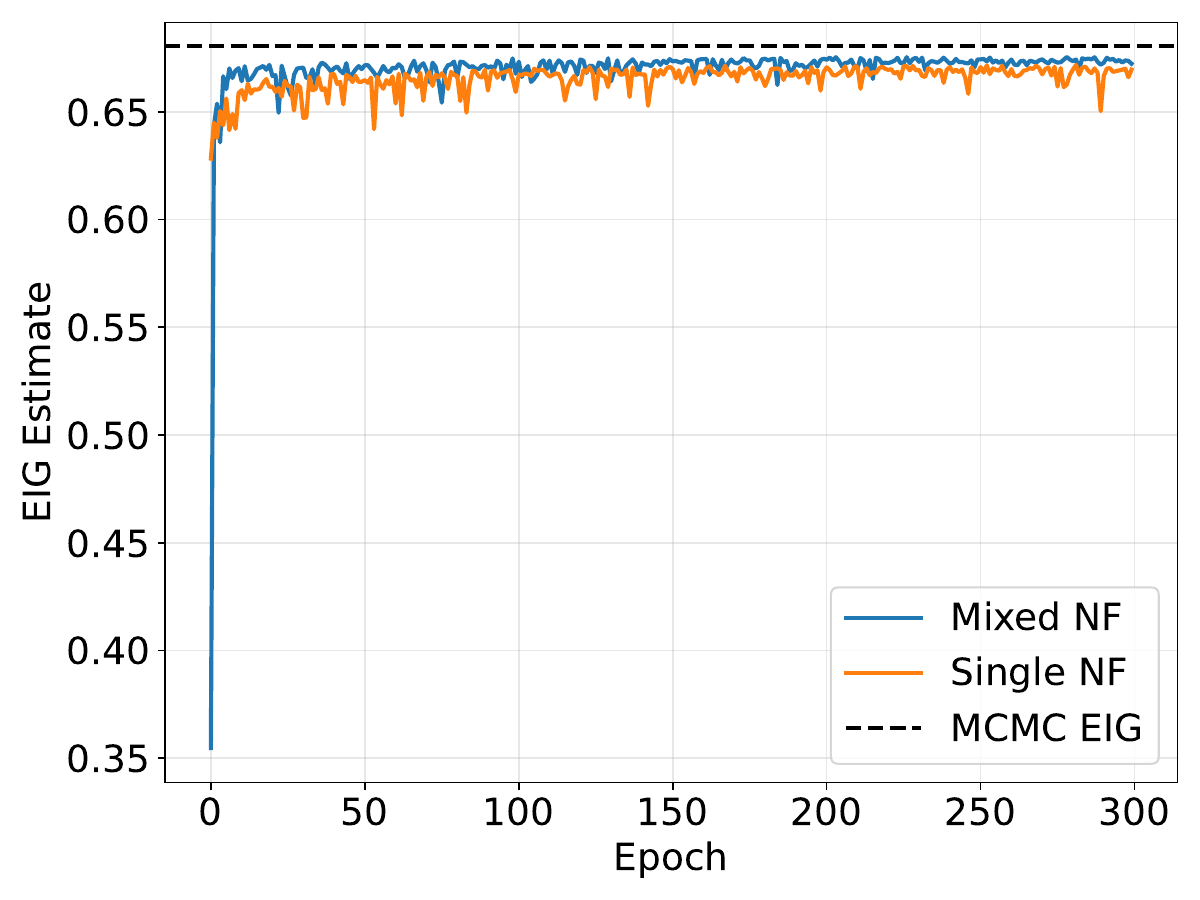}
    }
    \caption{
        Training trajectories of the variational EIG criterion at selected wave-group designs.
        The upper row corresponds to $r_s=0.25\,\mathrm{rad}$, for which the QoI law contains an atom at zero.
        The lower row corresponds to $r_s=0.1\,\mathrm{rad}$, for which the sampled QoI law is effectively continuous.
        Horizontal reference values are computed using the non-variational MCMC EIG estimator.
    }
    \label{fig:nf_training_comparison}
\end{figure*}

Figure~\ref{fig:nf_training_comparison} compares the training trajectories of the purely continuous NF and mixed NF formulations at selected designs under both thresholds.
The variational lower-bound estimates use the same sample size $N=2\times 10^4$ as in the toy example.
The reference EIG values are computed using a non-variational Markov chain Monte Carlo (MCMC)-based estimator of Zhong et al.~\cite{Zhong2026}.

For $r_s=0.25\,\mathrm{rad}$, the purely continuous NF produces unstable empirical trajectories that may exceed the reference EIG because zero-valued samples are evaluated using continuous density values rather than probability masses.
As discussed in Section~\ref{s:methods}, the resulting empirical criterion is not an estimator of the measure-theoretically valid Barber--Agakov lower bound.
The mixed NF instead assigns probability mass to the zero outcome and represents the positive component using a conditional NF, producing more stable EIG lower-bound estimates.

For $r_s=0.1\,\mathrm{rad}$, the sampled QoI law is effectively continuous, and the purely continuous NF is compatible with the observed distributional structure.
The mixed NF remains applicable, but its atom-probability component is unnecessary when the atom probability is zero and may introduce additional finite-sample or optimization variation.
These results are consistent with the measure-theoretic analysis in Section~\ref{s:methods}:
a mixed variational formulation is required when the QoI law contains an atom, whereas a purely continuous family is admissible when the target law is continuous.

\paragraph{EIG landscapes.}

\begin{figure*}[pos=t]
    \centering
    \subfloat[$r_s=0.25\,\mathrm{rad}$.]{
        \includegraphics[width=0.45\textwidth]{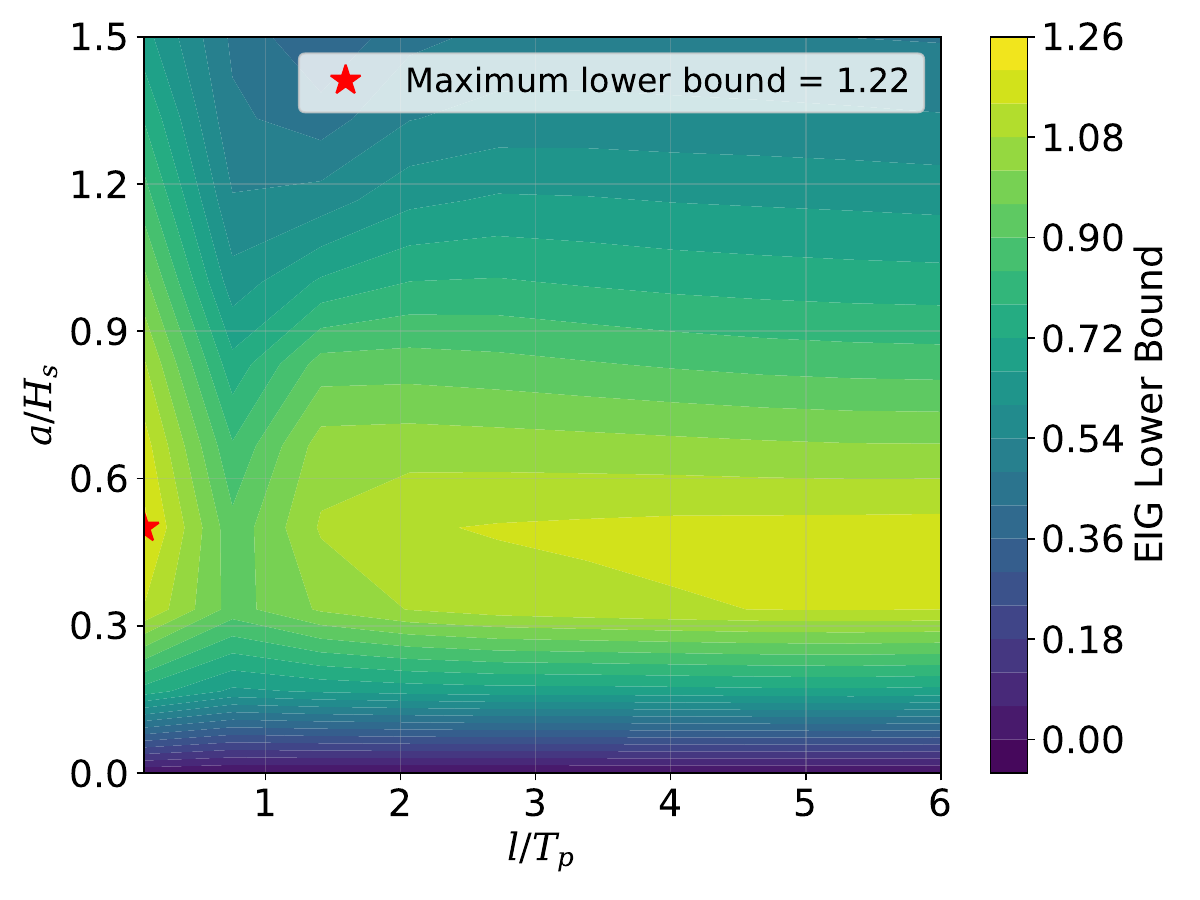}
    }
    \hspace{1em}
    \subfloat[$r_s=0.1\,\mathrm{rad}$.]{
        \includegraphics[width=0.45\textwidth]{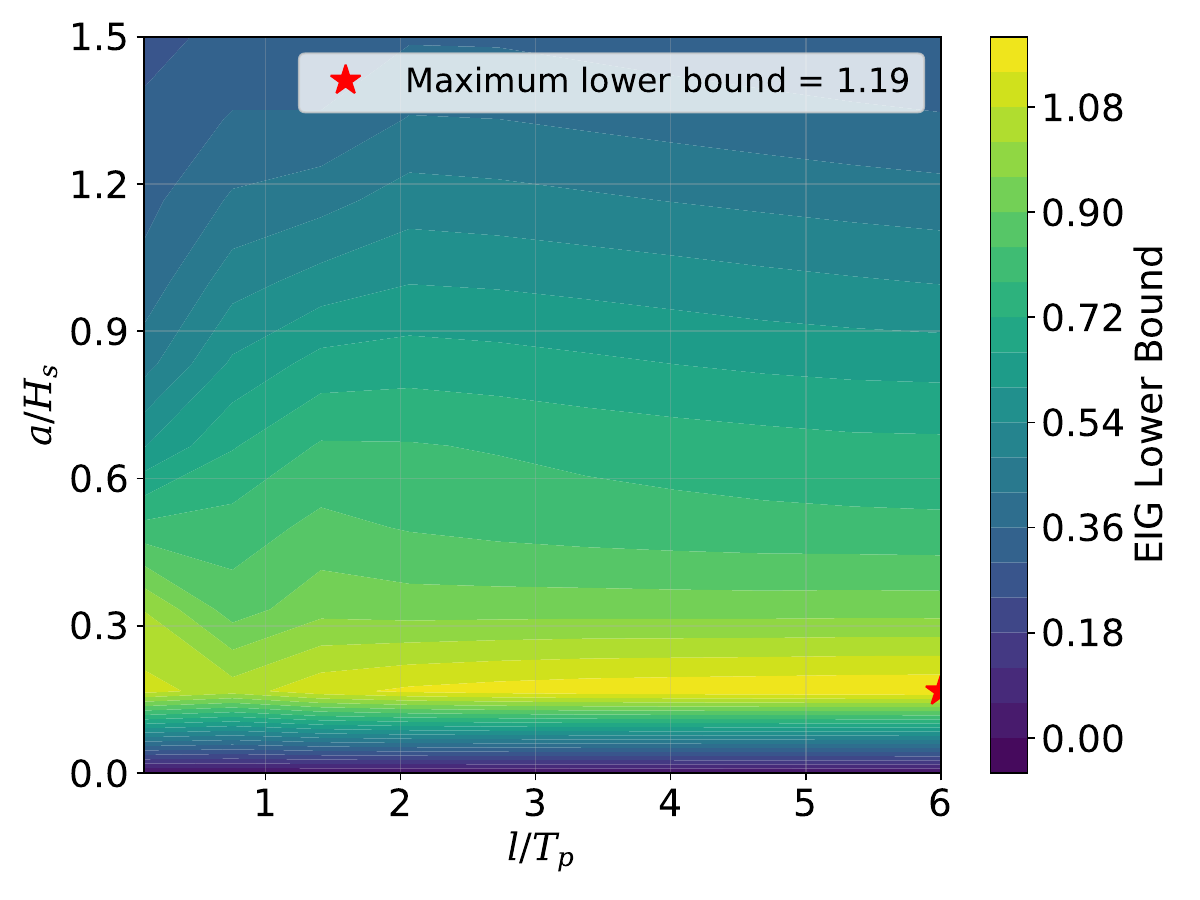}
    }
    \caption{
        Mixed-NF EIG lower-bound estimates over the wave-group design space.
        The red star denotes the maximizing grid point.
    }
    \label{fig:utility_lb_comparison}
\end{figure*}

Figure~\ref{fig:utility_lb_comparison} shows the mixed-NF EIG lower-bound estimates over the gridded wave-group design space.
The mixed formulation is used for both thresholds so that the same variational framework can be applied to the mixed and effectively continuous cases.

For $r_s=0.25\,\mathrm{rad}$, the EIG lower-bound landscape contains a localized region of larger values.
For $r_s=0.1\,\mathrm{rad}$, the estimated EIG varies more gradually across the design space.
The maximizing designs differ between thresholds, demonstrating that the wave group expected to be most informative about the safety QoI depends on the roll-angle threshold defining that QoI.

The plotted values are variational lower-bound estimates rather than exact EIG values.
They are therefore interpreted primarily through comparisons across designs.
A larger value indicates a larger estimated EIG under the fitted variational approximation, rather than an exact numerical value of the mutual information.

\paragraph{Posterior diagnostics at selected designs.}

\begin{figure*}[pos=t]
    \centering
    \subfloat[Highest-EIG design, realization 1.]{
        \includegraphics[width=0.48\linewidth]{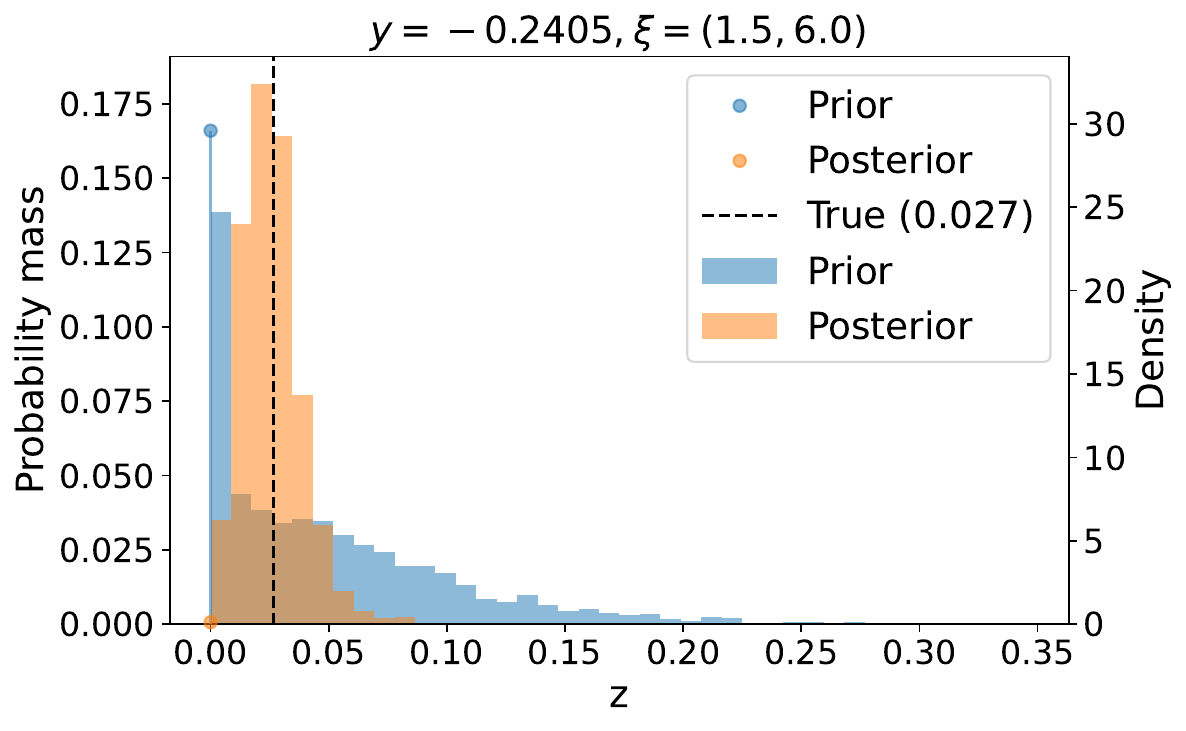}
    }
    \hfill
    \subfloat[Low-EIG reference design, realization 1.]{
        \includegraphics[width=0.48\linewidth]{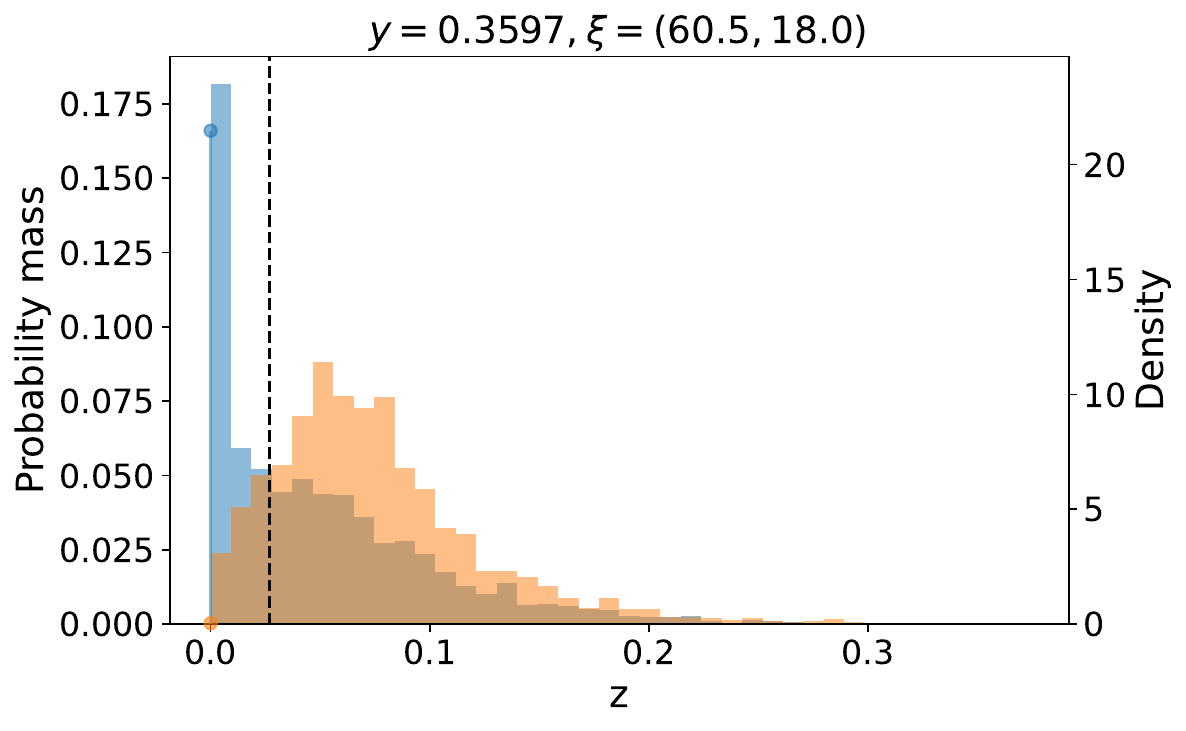}
    }\\
    \subfloat[Highest-EIG design, realization 2.]{
        \includegraphics[width=0.48\linewidth]{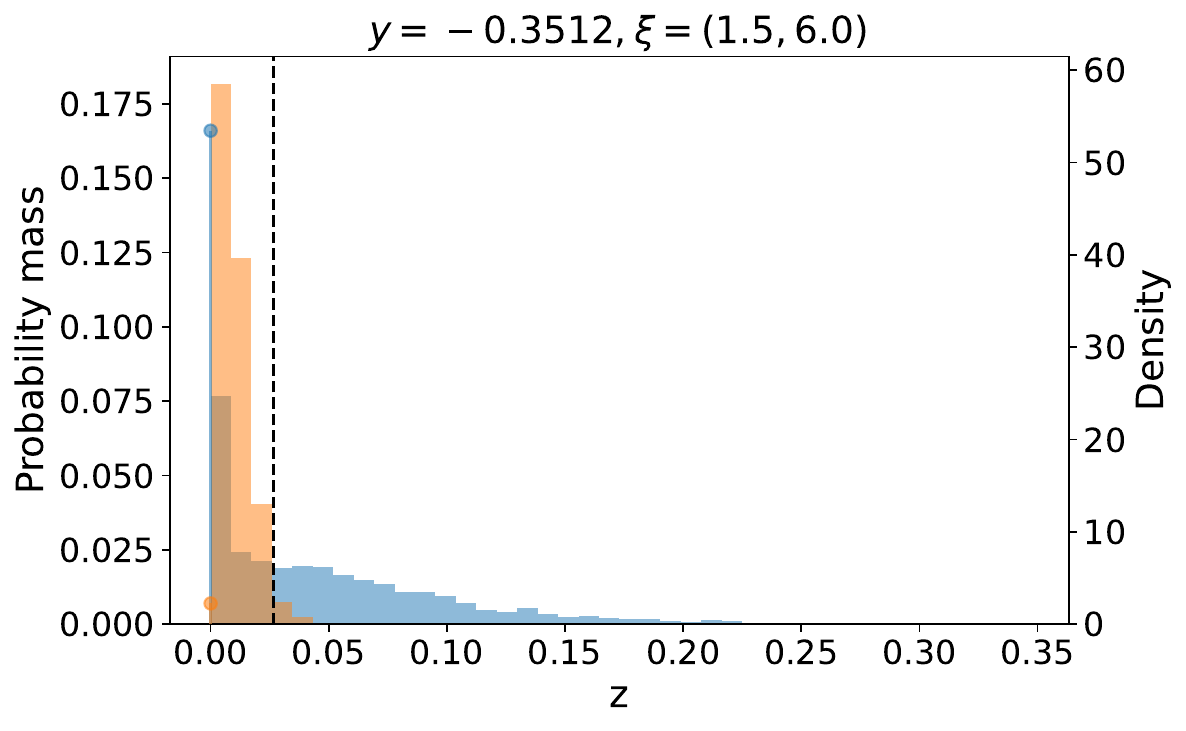}
    }
    \hfill
    \subfloat[Low-EIG reference design, realization 2.]{
        \includegraphics[width=0.48\linewidth]{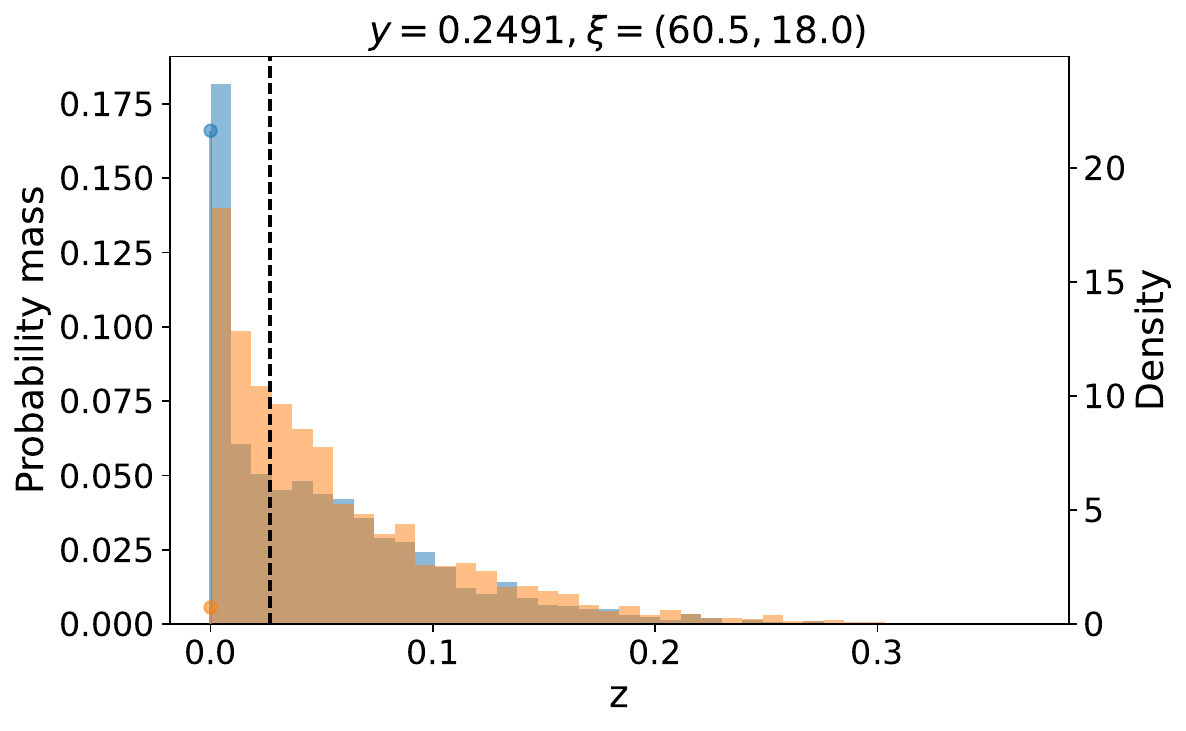}
    }
    \caption{
        Prior and posterior laws of the safety QoI for $r_s=0.25\,\mathrm{rad}$.
        Each panel displays the prior law, variational posterior law, and true QoI value.
    }
    \label{fig:prior_posterior_Z_GO_r0p25}
\end{figure*}

\begin{figure*}[pos=t]
    \centering
    \subfloat[Highest-EIG design, realization 1.]{
        \includegraphics[width=0.48\linewidth]{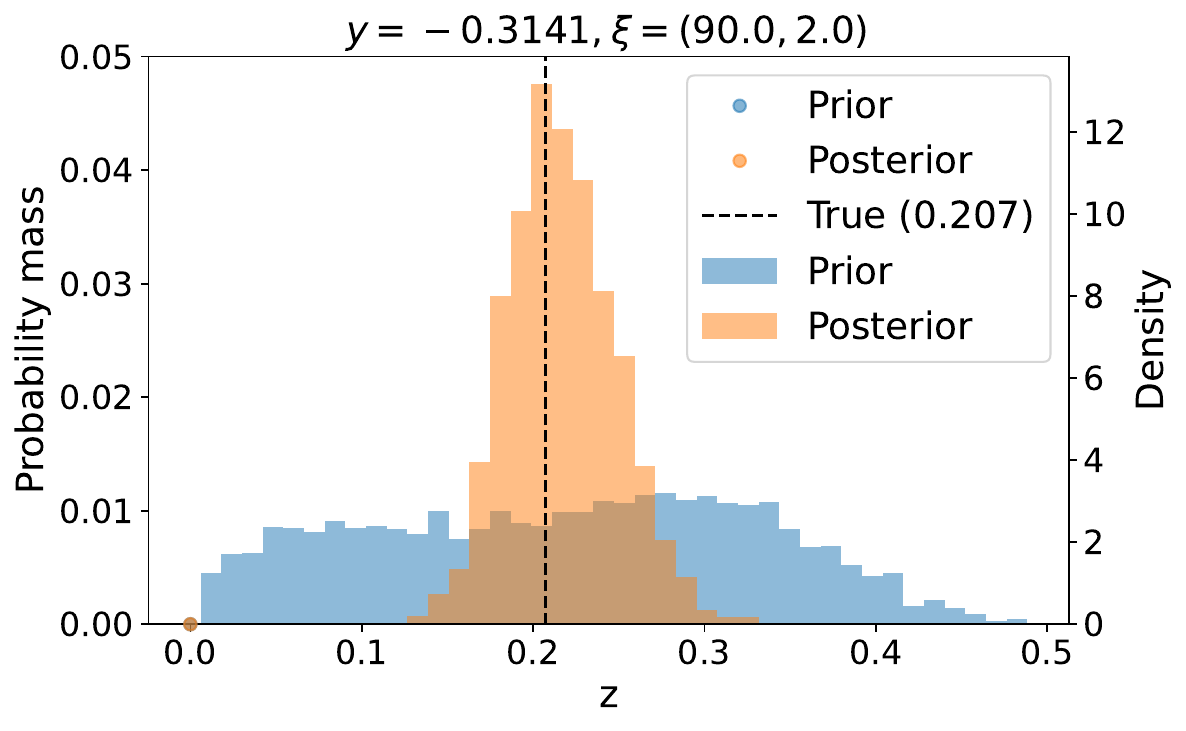}
    }
    \hfill
    \subfloat[Low-EIG reference design, realization 1.]{
        \includegraphics[width=0.48\linewidth]{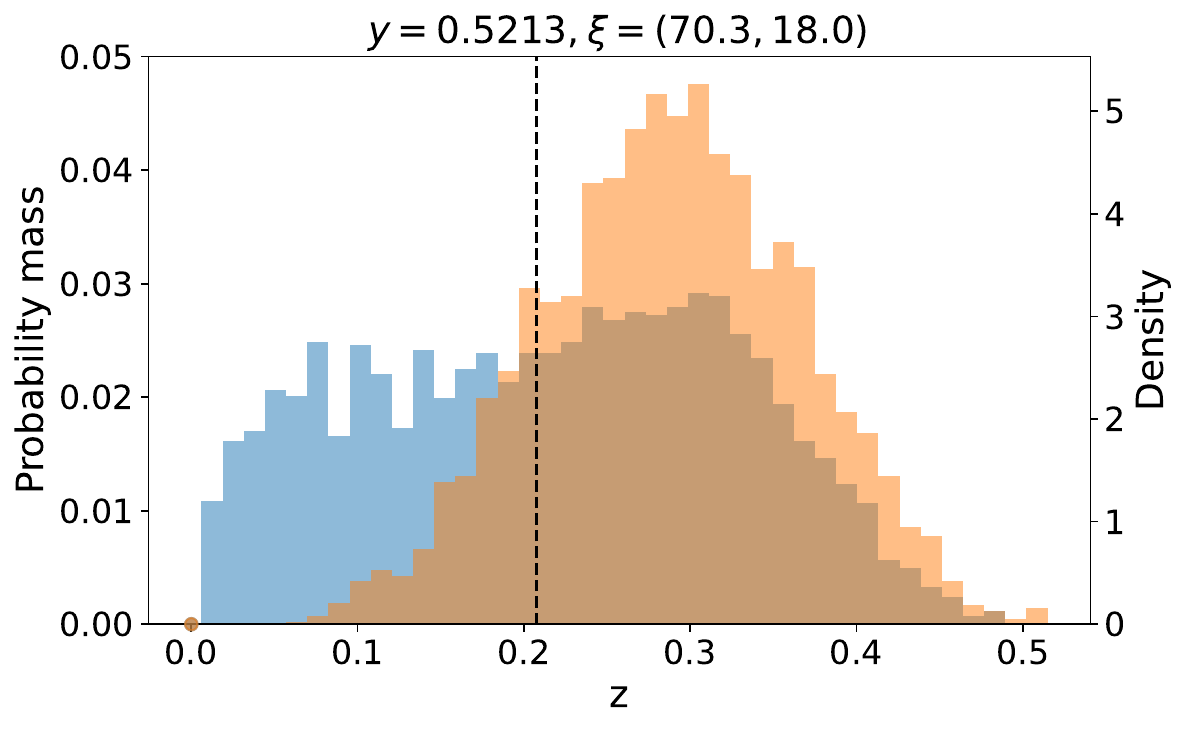}
    }\\
    \subfloat[Highest-EIG design, realization 2.]{
        \includegraphics[width=0.48\linewidth]{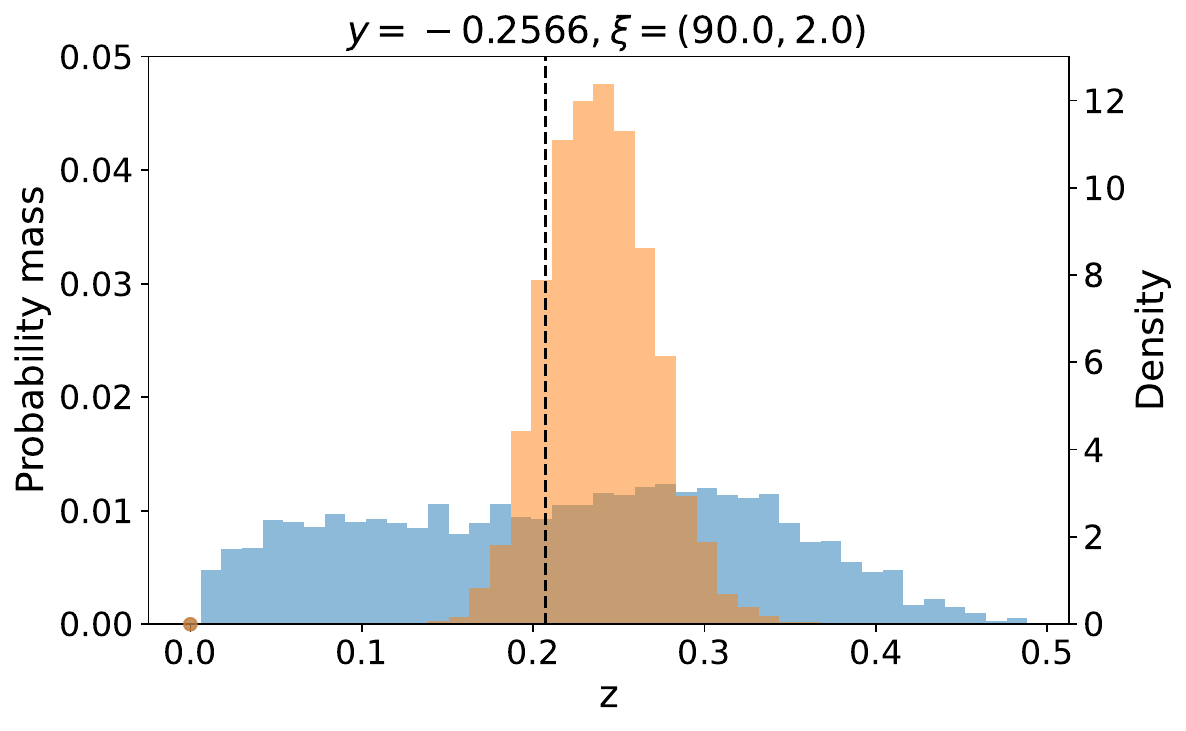}
    }
    \hfill
    \subfloat[Low-EIG reference design, realization 2.]{
        \includegraphics[width=0.48\linewidth]{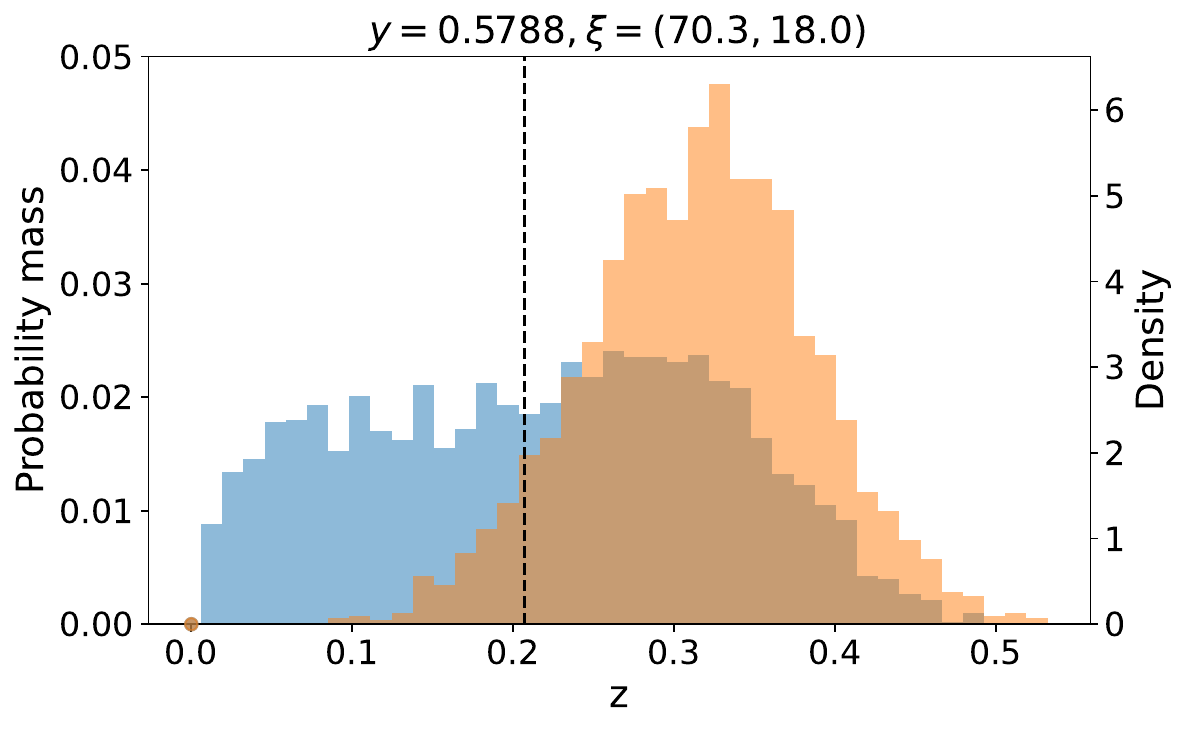}
    }
    \caption{
        Prior and posterior laws of the safety QoI for $r_s=0.1\,\mathrm{rad}$.
        Each panel displays the prior law, variational posterior law, and true QoI value.
    }
    \label{fig:prior_posterior_Z_GO_r0p1}
\end{figure*}

Figures~\ref{fig:prior_posterior_Z_GO_r0p25} and~\ref{fig:prior_posterior_Z_GO_r0p1} compare posterior QoI laws obtained at the highest-EIG design and at a low-EIG reference design.
Each row corresponds to an independently generated observation at a fixed true parameter realization sampled from the prior.
These examples are posterior diagnostics rather than estimates of EIG.
They illustrate the inference outcomes produced by the selected and reference designs for representative observations.

For $r_s=0.25\,\mathrm{rad}$, the highest-EIG design produces posterior laws that are more concentrated near the true QoI value in the displayed cases.
The low-EIG reference design produces broader or more displaced posterior laws.
The results for $r_s=0.1\,\mathrm{rad}$ show the same qualitative distinction in the effectively continuous regime.

The displayed realizations do not establish superiority for every possible observation.
The GO-OED criterion averages over the prior predictive law of $Y$, so the design comparison is based on expected information.
The posterior examples provide representative illustrations of the inference behavior underlying the EIG criterion.

\section{Conclusions}
\label{s:conclusion}

This work developed a variational GO-OED formulation for mixed discrete--continuous QoIs arising in probabilistic mechanics.
The motivating application is a threshold-based ship roll safety metric that equals zero when no threshold exceedance occurs and varies continuously over positive values otherwise.
Whenever the non-exceedance regime has positive probability, the induced QoI law contains an atom at zero together with a continuous component.
This structure determines the appropriate dominating measure for the posterior QoI law and its variational approximation.
If the posterior law contains an atom, a purely continuous variational law does not dominate it.
The corresponding Kullback--Leibler divergence is infinite, so the correctly defined Barber--Agakov objective reduces to the trivial lower bound $-\infty$.
A finite implementation that instead evaluates atom samples using continuous density values changes the objective and is not an estimator of the valid lower bound.

The proposed mixed variational formulation resolves this incompatibility by modeling the conditional atom probability and continuous component separately.
Samples at the atom are scored using probability mass, while samples in the continuous component are scored using density.
Under the stated support conditions, the resulting approximation yields a finite Barber--Agakov lower bound.
The analytical example isolates the measure mismatch in a setting where the exact goal-oriented EIG is available.
The mixed approximation reproduces the correct EIG landscape, whereas the purely continuous implementation can produce unstable empirical scores and values above the exact EIG when probability masses are replaced by density values.

The ship roll application demonstrates the methodology for a nonlinear stochastic dynamical system.
A localized wave-group representation provides a physically meaningful experimental design space, while a threshold-specific surrogate observation map enables repeated evaluation of the roll response and temporal exceedance QoI.
For the higher roll-angle threshold, the QoI law contains an atom at zero, and the mixed NF produces stable EIG lower-bound estimates and interpretable design landscapes.
For the lower threshold, no atom is observed in the sampled prior ensemble, and the QoI law is treated as effectively continuous.
In this regime, a purely continuous NF is compatible with the observed distributional structure, while the mixed formulation remains applicable as the estimated atom probability approaches zero.
The results show that the probabilistic structure induced by the response metric determines the appropriate variational family and scoring measure in GO-OED.

The same issue can arise broadly in probabilistic mechanics when thresholding, censoring, clipping, or event-based transformations map a positive-probability set of uncertain inputs to a common QoI value while other inputs produce continuously varying responses.
Representative examples include failure severity conditional on failure occurrence, exceedance duration, accumulated damage beyond onset, structural uptime and downtime fractions, and reliability metrics with a non-event state.
The analysis provides a diagnostic for such problems.
A posterior QoI law containing an atom requires a variational family that assigns probability mass to that outcome in order to obtain a finite-gap Barber--Agakov approximation.

Several limitations motivate further development.
The ship application relies on a surrogate whose largest errors occur for short, high-amplitude wave groups, where the roll dynamics become strongly nonlinear.
Adaptive sampling, multifidelity surrogates, and physics-informed approximations could improve accuracy in this region and quantify the sensitivity of the selected design to surrogate error.
The present study considers nonsequential design, while sequential formulations would allow experimental conditions to adapt as observations are collected.
The wave-group representation is tailored to a narrow-band sea state, and broader-band or non-Gaussian environments may require different reduced parameterizations and design spaces.
The current formulation also considers a scalar QoI with a single atom and an otherwise absolutely continuous law.
Extensions to multiple atoms, mixed categorical--continuous structures, and higher-dimensional mixed QoIs provide natural directions for future work in probabilistic design and inference.
\section*{Acknowledgments}

This work relates to Department of Navy award N00014-24-1-2266 and N000142512411 issued by the
Office of Naval Research.
This research was supported in part through computational resources and services provided by Advanced Research Computing at the University of Michigan, Ann Arbor.

\appendix

\section{Derivation of the Exact EIG for the Analytical Example}
\label{app:analytical_mixed_example}

This appendix derives the exact goal-oriented EIG for the validation problem introduced in Section~\ref{ss:toy_problem}.
Let $\varphi$ and $\Phi$ denote the standard normal density and cumulative distribution function, respectively.
The QoI is
\begin{align}
    Z
    =
    \max\{0,\Param_1-c\},
\end{align}
where $\Param_1\sim\mathcal{N}(0,1)$.
Realizations satisfying $\param_1\leq c$ are mapped to the common value $z_0=0$, whereas realizations satisfying $\param_1>c$ yield $z=\param_1-c>0$.
The prior law of $Z$ is therefore
\begin{align}
    \mathbb{P}_Z(\mathrm{d}z)
    =
    \Phi(c)\,
    \delta_0(\mathrm{d}z)
    +
    \varphi(z+c)\,
    \mathbf{1}_{(0,\infty)}(z)\,
    \mathrm{d}z.
    \label{e:toy_prior_mixed}
\end{align}
The prior atom probability is
\begin{align}
    \pi_0
    =
    \mathbb{P}(Z=0)
    =
    \Phi(c).
\end{align}
The normalized density conditional on membership in the positive continuous component is
\begin{align}
    f_{\mathrm{c}}(z)
    =
    \frac{
        \varphi(z+c)
    }{
        1-\Phi(c)
    },
    \qquad
    z>0.
    \label{e:toy_prior_positive_density}
\end{align}

The observation model is
\begin{align}
    Y
    =
    g(\design)^\top\Param+\Epsilon,
    \qquad
    g(\design)
    =
    \begin{bmatrix}
        \cos\design\\
        \sin\design
    \end{bmatrix},
    \qquad
    \Epsilon
    \sim
    \mathcal{N}(0,\sigma_\epsilon^2),
\end{align}
with $\design\in\designset=[0,\pi]$.
Since $\|g(\design)\|_2=1$, the prior predictive law of the observation is
\begin{align}
    Y|\design
    \sim
    \mathcal{N}
    \left(
        0,
        1+\sigma_\epsilon^2
    \right).
    \label{e:toy_observation_predictive}
\end{align}

The pair $(\Param_1,Y)$ is jointly Gaussian, with
\begin{align}
    \operatorname{Cov}(\Param_1,Y|\design)
    =
    \cos\design.
\end{align}
The posterior marginal law of the threshold-driving parameter is consequently
\begin{align}
    \Param_1|Y=y,\design
    \sim
    \mathcal{N}
    \left(
        m(y,\design),
        s^2(\design)
    \right),
    \label{e:toy_theta1_posterior}
\end{align}
where
\begin{align}
    m(y,\design)
    &=
    \frac{
        \cos\design
    }{
        1+\sigma_\epsilon^2
    }
    y,
    \label{e:toy_posterior_mean}\\
    s^2(\design)
    &=
    1-
    \frac{
        \cos^2\design
    }{
        1+\sigma_\epsilon^2
    }.
    \label{e:toy_posterior_variance}
\end{align}

The posterior atom probability is
\begin{align}
    \pi_0(y,\design)
    &=
    \mathbb{P}
    \left(
        Z=0
        |
        Y=y,\design
    \right)
    \nonumber\\
    &=
    \mathbb{P}
    \left(
        \Param_1\leq c
        |
        Y=y,\design
    \right)
    \nonumber\\
    &=
    \Phi
    \left(
        \frac{
            c-m(y,\design)
        }{
            s(\design)
        }
    \right).
    \label{e:toy_posterior_atom}
\end{align}
For $z>0$, the transformation $z=\param_1-c$ gives the unnormalized posterior density on the positive component,
\begin{align}
    \widetilde{f}_{\mathrm{c}}(z|y,\design)
    =
    \frac{1}{
        s(\design)
    }
    \varphi
    \left(
        \frac{
            z+c-m(y,\design)
        }{
            s(\design)
        }
    \right).
    \label{e:toy_posterior_continuous_component}
\end{align}
Its integral equals the posterior probability of the continuous component:
\begin{align}
    \int_0^\infty
    \widetilde{f}_{\mathrm{c}}(z|y,\design)
    \,\mathrm{d}z
    =
    1-\pi_0(y,\design).
\end{align}
The corresponding normalized conditional density is
\begin{align}
    f_{\mathrm{c}}(z|y,\design)
    =
    \frac{
        \widetilde{f}_{\mathrm{c}}(z|y,\design)
    }{
        1-\pi_0(y,\design)
    },
    \qquad
    z>0.
    \label{e:toy_posterior_positive_density}
\end{align}

The posterior law of $Z$ can therefore be written as
\begin{align}
    \mathbb{P}_{Z|Y,\design}(\mathrm{d}z|y,\design)
    &=
    \pi_0(y,\design)\,
    \delta_0(\mathrm{d}z)
    \nonumber\\
    &\quad+
    \left[
        1-\pi_0(y,\design)
    \right]
    f_{\mathrm{c}}(z|y,\design)\,
    \mathbf{1}_{(0,\infty)}(z)\,
    \mathrm{d}z.
    \label{e:toy_posterior_mixed}
\end{align}

Let
\begin{align}
    \nu
    =
    \delta_0+\mu,
\end{align}
where $\mu$ denotes Lebesgue measure on $(0,\infty)$.
The Radon--Nikodym derivatives of the prior and posterior laws with respect to $\nu$ are
\begin{align}
    p_{Z,\nu}(z)
    &=
    \begin{cases}
        \Phi(c),
        & z=0,\\[0.3em]
        \varphi(z+c),
        & z>0,
    \end{cases}
    \label{e:toy_prior_RN}\\
    p_{Z|Y,\design,\nu}(z|y,\design)
    &=
    \begin{cases}
        \pi_0(y,\design),
        & z=0,\\[0.3em]
        \widetilde{f}_{\mathrm{c}}(z|y,\design),
        & z>0.
    \end{cases}
    \label{e:toy_posterior_RN}
\end{align}

For a realized observation $y$, the realized information gain is
\begin{align}
    &D_{\mathrm{KL}}
    \left(
        \mathbb{P}_{Z|Y,\design}(\cdot|y,\design)
        \,||\,
        \mathbb{P}_Z
    \right)
    \nonumber\\
    &\quad=
    \pi_0(y,\design)
    \log
    \frac{
        \pi_0(y,\design)
    }{
        \Phi(c)
    }
    \nonumber\\
    &\qquad+
    \int_0^\infty
    \widetilde{f}_{\mathrm{c}}(z|y,\design)
    \log
    \frac{
        \widetilde{f}_{\mathrm{c}}(z|y,\design)
    }{
        \varphi(z+c)
    }
    \,\mathrm{d}z.
    \label{e:toy_exact_kl}
\end{align}
The first term accounts for the change in atom probability, while the second accounts for the change in the continuous part of the law.

The same realized information gain can be expressed using the normalized component densities:
\begin{align}
    &D_{\mathrm{KL}}
    \left(
        \mathbb{P}_{Z|Y,\design}(\cdot|y,\design)
        \,||\,
        \mathbb{P}_Z
    \right)
    \nonumber\\
    &\quad=
    D_{\mathrm{KL}}
    \left(
        \operatorname{Bern}
        \left(
            \pi_0(y,\design)
        \right)
        \,||\,
        \operatorname{Bern}
        \left(
            \Phi(c)
        \right)
    \right)
    \nonumber\\
    &\qquad+
    \left[
        1-\pi_0(y,\design)
    \right]
    D_{\mathrm{KL}}
    \left(
        f_{\mathrm{c}}(\cdot|y,\design)
        \,||\,
        f_{\mathrm{c}}
    \right).
    \label{e:toy_exact_kl_decomposition}
\end{align}

The exact goal-oriented EIG is obtained by averaging the realized information gain over the prior predictive law of $Y$:
\begin{align}
    U_{\mathrm{GO}}(\design)
    &=
    \mathcal{I}(Z;Y|\design)
    \nonumber\\
    &=
    \int
    D_{\mathrm{KL}}
    \left(
        \mathbb{P}_{Z|Y,\design}(\cdot|y,\design)
        \,||\,
        \mathbb{P}_Z
    \right)
    p_{Y|\design}(y|\design)
    \,\mathrm{d}y,
    \label{e:toy_exact_utility}
\end{align}
where
\begin{align}
    p_{Y|\design}(y|\design)
    =
    \frac{1}{
        \sqrt{
            2\pi
            \left(
                1+\sigma_\epsilon^2
            \right)
        }
    }
    \exp
    \left[
        -
        \frac{
            y^2
        }{
            2
            \left(
                1+\sigma_\epsilon^2
            \right)
        }
    \right].
\end{align}
Equations~\eqref{e:toy_exact_kl} and~\eqref{e:toy_exact_utility} provide an exact integral representation of the EIG.
The resulting nested one-dimensional integrals over $z$ and $y$ can be evaluated to arbitrary numerical accuracy using quadrature.

The design dependence follows directly from
\eqref{e:toy_posterior_mean}--\eqref{e:toy_posterior_variance}.
At $\design=\pi/2$, $\cos\design=0$, and
\begin{align}
    m(y,\pi/2)
    =
    0,
    \qquad
    s^2(\pi/2)
    =
    1.
\end{align}
The posterior law of $\Param_1$, and hence that of $Z$, then coincides with its prior law for every observation.
Consequently,
\begin{align}
    U_{\mathrm{GO}}(\pi/2)
    =
    0.
\end{align}

For $\cos\design\neq 0$, the observation can be rescaled as
\begin{align}
    \frac{Y}{\cos\design}
    =
    \Param_1
    +
    \frac{
        \sin\design\,\Param_2+\Epsilon
    }{
        \cos\design
    }.
\end{align}
The effective noise is Gaussian and independent of $\Param_1$, with variance
\begin{align}
    \sigma_{\mathrm{eff}}^2(\design)
    =
    \frac{
        \sin^2\design+\sigma_\epsilon^2
    }{
        \cos^2\design
    }
    =
    \frac{
        1-\cos^2\design+\sigma_\epsilon^2
    }{
        \cos^2\design
    }.
    \label{e:toy_effective_noise}
\end{align}
This variance decreases as $|\cos\design|$ increases.
An observation with larger effective noise can be obtained from one with smaller effective noise by adding independent Gaussian noise.
The data-processing inequality therefore implies that
$\mathcal{I}(Z;Y|\design)$ is nondecreasing in $|\cos\design|$.
The EIG is consequently maximized when
$|\cos\design|=1$, corresponding to $\design^\ast
    \in
    \{0,\pi\}$.

\printcredits
 
\section*{Declaration of Generative AI and AI-Assisted Technologies in the Writing Process}

During the preparation of this work, the authors used ChatGPT to assist with language editing, organization, clarity, and consistency checks of mathematical notation and presentation. The authors reviewed and revised all AI-assisted text and take full responsibility for the content of the manuscript.

\bibliographystyle{cas-model2-names}
\bibliography{references}

\end{document}